\documentclass[12pt, preprint]{aastex}

\newcommand\um{\ifmmode{\mu{\rm m}}\else{$\mu$m}\fi}

\def\eq#1{\begin{equation} #1 \end{equation}}

\begin{document}
\title{Silicates in Ultra-Luminous Infrared Galaxies
}
\author{
M. M. Sirocky\altaffilmark{1}, 
N. A. Levenson\altaffilmark{1}, 
M. Elitzur\altaffilmark{1}, 
H. W. W. Spoon\altaffilmark{2},
and
L. Armus\altaffilmark{3}}
\altaffiltext{1}{Department of Physics and Astronomy, University of Kentucky,
Lexington, KY 40506; 
sirockmm@pa.uky.edu,
levenson@pa.uky.edu,
moshe@pa.uky.edu}
\altaffiltext{2}{Astronomy Department, Cornell University,
Ithaca, NY 14853;
spoon@isc.astro.cornell.edu}
\altaffiltext{3}{\textit{Spitzer} Science Center, California Institute of Technology, Pasadena, CA 91125; lee@ipac.caltech.edu}
\begin{abstract}
We analyze the mid-infrared (MIR) spectra of 
ultraluminous infrared galaxies (ULIRGs) observed with
the \textit{Spitzer Space Telescope}'s Infrared Spectrograph.
Dust emission dominates the MIR spectra of ULIRGs, 
and the reprocessed radiation that emerges is independent
of the underlying heating spectrum.
Instead, the resulting emission depends sensitively on
the geometric distribution of the dust, 
 which we diagnose with comparisons of 
numerical simulations of radiative transfer. 
Quantifying the silicate emission and absorption features 
that appear near 10 and 18\um{} requires
a reliable determination  of the continuum, and we demonstrate that
including a  measurement of the  continuum
at intermediate wavelength (between the features) produces 
accurate results at all optical depths. 
With high-quality spectra, we successfully 
use the silicate features 
to constrain the dust chemistry. 
The observations of the ULIRGs and local sightlines 
require dust that has a
relatively high 18/10\um{} absorption ratio of the silicate features 
(around 0.5).
Specifically, the cold dust of 
Ossenkopf et al. (1992) 
is consistent with the observations, while
other dust models are not. 
We use the silicate feature strengths to identify two families of 
ULIRGs, in which the dust distributions are fundamentally different.
Optical spectral classifications are related to these families.
In ULIRGs that harbor an active galactic nucleus,
the spectrally  broad lines are detected only
when the nuclear surroundings are clumpy.
In contrast, the sources of lower ionization optical
spectra are 
deeply embedded in smooth distributions of optically thick dust.
\end{abstract}

\section{Introduction}

High infrared luminosity ($L_{IR} > 10^{12} L_\sun$) characterizes
ultraluminous infrared galaxies (ULIRGs).
The underlying energy source may be accretion onto a supermassive black
hole in an active galactic nucleus (AGN), intense bursts of star formation,
or a combination of the two \citep{San88a,Gen98}.
Because the total luminosities are large, ULIRGs are critical sites
to consider in obtaining a complete account
of star formation, black hole  growth, and the relationship between these 
two phenomena over cosmic time.
In all cases, dust is responsible for reprocessing the
intrinsic hard radiation to emerge at longer wavelengths.
This dust reprocessing erases the spectral signatures that would
reveal the nature of the original source, but the resulting spectra
serve as probes of the dust itself.

While most of the dust emission is continuum radiation,
in the mid-infrared (MIR), two of its spectral features
are observable,  arising near 10 and 18\um.
These are attributed to silicates, 
especially amorphous analogs of pyroxenes and olivines 
\citep[and references therein]{Dra03r}. 
Both features have been measured in stellar observations 
for decades \citep*{Gil68,Woo69,Low70}. 
Observations of the stronger 10\um{} absorption feature 
in starburst galaxies and AGNs 
have a similarly long history \citep*[\textit{e.g.},][]{Gil75,Rie75a,Rie75b,Kle76},
but only recently has the high sensitivity of instruments
in space made measurement of 
the 18\um{} feature 
in large numbers of galaxies feasible.

Detecting silicate features in emission  can identify the
dusty medium as optically thin, and deep absorption characterizes 
optically thick obscuration. 
In detail, however, 
quantifying the optical depth requires a complete radiative transfer
calculation because the same material produces both the continuum
and line features.
Thus, the apparent optical depth that silicate absorption strength yields
is not the true optical depth along the line of sight, although it is
frequently treated as such a direct measurement \citep[\textit{e.g.},][]{Shi06,Gal04,Mai01}.  
Compared with complete calculations, however,
high signal-to-noise spectra of both of these features together
effectively reveal the geometric distribution of the dust, including
the total line-of-sight optical depth. 
Moreover, while the continuum emission depends only weakly on
the dust chemistry, comparison of the two silicate features
provides one of the few accessible diagnostics of the dust composition.

Infrared studies that aim to deduce the nature of sources and surrounding dust geometry 
suffer from the degeneracy of the
problem: a  large range of different configurations
 produces very similar overall spectral energy distributions.
Reaching conclusions about geometry, let alone dust chemistry,
based on the spectral shape
is difficult if not impossible \citep[\textit{e.g.}][]{Vin03}.
However, extremely deep  10\um{} silicate absorption
proves to be a good discriminant of clumpy and smooth dust distributions, 
as we show in \citet{Lev07}.
Here we extend this analysis to include both the 10 and 18\um{}
silicate features.
We apply 
 detailed models to
MIR spectral measurements of ULIRGs 
to probe the 
dust chemistry and geometric distribution of material around the
galaxies' powerful nuclear sources.
While the MIR spectra of ULIRGs have been studied previously
\citep[\textit{e.g.},][]{Arm04,Arm06,Arm07,Des07,Far07,Mar07,Spo04,Spo06}, 
here we investigate both 
silicate features in detail.

We consider the ULIRGs contained in the samples
of \citet{Spo07} and \citet{Ima07}, which were all
observed using the 
Infrared Spectrograph (IRS; \citealt{Hou04}) on board
the \textit{Spitzer Space Telescope} \citep{Wer04}.
All the observations were obtained in low-resolution
mode, which provides spectral resolution $R\sim 100$
and complete coverage from 5--35\um.
The slits are up to 3.7 and 10.7\arcsec{} wide
in the short- and long-wavelength modules, respectively.
Although these slits cover large areas of the galaxies
(which have redshifts up to 0.93), 
the concentrated emission dominates the
flux of the nuclear spectra we discuss.
See \citet{Spo07} for additional information about 
the data reduction.

\section{Modeling Dust Reprocessing} 

Dust reprocesses the heating spectrum, and the reprocessed radiation emerges at IR
wavelengths.  In contrast to standard extinction along the line of
sight, whereby dust modifies the underlying spectrum of the source to
produce the transmitted spectrum, this reprocessed dust emission
fundamentally erases the spectral signature of the heating source.
The emitting dust need not be located along the line of sight to the
source. 
In the absence of other obscuration, both
the intrinsic source and the reprocessed dust emission may then be
detected, as is the case, for example, in
type 1 AGNs.  However, because the stellar and AGN engines of ULIRGs are
intrinsically weak in the IR regime, in all cases the reprocessed dust
emission dominates the resulting IR spectrum. 

We calculate the radiative transfer through the dusty medium
 using the one-dimensional 
code DUSTY \citep{Ive99}. 
The emergent spectrum depends on 
the dust composition,
the geometric distribution
of the dust, and
the initial heating spectrum.
We adopt the grain size distribution of \citet*{Mat77}.
Variations of the grain sizes affect only
NIR and shorter wavelength emission and
do not alter silicate emission or absorption
at 10 and 18\um, as long as the grains sizes do not exceed about 1\um.
The grain composition is
53\% silicate and 47\% graphite.
We consider the optical properties of three different 
models of dust chemistry: from \citet*{Oss92},  
cool, oxygen-rich silicates that may be typical of the diffuse interstellar
medium (``OHMc'') and 
warm, oxygen-deficient silicates
that may be typical of circumstellar regions (``OHMw''), and
``astronomical silicate'' from 
\citet[``Draine'']{Dra03d1,Dra03d2}. 
In all cases, we use the graphite optical properties of \citet{Dra03d1,Dra03d2}.
We explore further variations in composition 
below.  We do not include any polycyclic aromatic
hydrocarbons (PAHs)
in the numerical models, and 
we restrict our detailed analysis below to sources
that exhibit weak PAH emission, in which the
silicate and graphite grain emission dominates the MIR spectra.

We examine 
three geometries of the dusty material:  
a normally-illuminated slab, 
a smooth spherical shell, and a clumpy
 spherical shell.
The spherical symmetry of the smooth and clumpy distributions 
minimizes the number of model variables. 
We use the formalism of \citet{Nen02} to calculate the
clumpy models. 
The maximum dust temperature is 1500K, at which dust sublimates.
The inner surface of the slab, shell, or
innermost clouds is located at the distance, $R_d$,
corresponding to this temperature.
This distance at which the dust sublimation
temperature is reached depends on the  luminosity
of the central source and its spectral shape.
For an incident AGN spectral shape, 
$R_d \sim 0.4 \sqrt{L_{45}}\mathrm{\, pc} $ for the typical grains, 
where $L_{45}$ is the bolometric luminosity in units of
$10^{45} \mathrm{\, erg\, s^{-1}}$. 
For all three geometries, we consider a  range of total optical depths 
$0.1 \leq \tau_V \leq 400$ to the source,
where $\tau_V$ is the optical depth at 0.55\um. 
The slab is geometrically thin, whereas
the radial density distribution and the total extent of the dust
can vary in the spherical models.
In the continuous shell, we 
consider dust density profiles proportional to $r^{-p}$, where
$p = 0$, 1, or 2, and the 
shell thickness, $Y = R/R_d$, ranges from 
1.25 to 1000.
The clumpy distribution similarly contains clouds 
whose number is proportional to $ r^{-p}$, located between $R_d$
and some maximum thickness, typically $30 R_d$.
The individual clouds are optically thick,
each with $\tau_{V,C}$ ranging from 10 to 80.
The number of clouds along radial rays is distributed stochastically according
to Poisson statistics around a prescribed mean value, which we
designate the mean number of clouds, $N_0$.  In our computations,
$1\le N_0\le 10$.  Thus, the average total optical depth along radial
rays, $N_0 \tau_{V,C}$, ranges from 10 to 800 through these clumpy
distributions.

\begin{figure}[htb!]
\centerline{\includegraphics[width=3in]{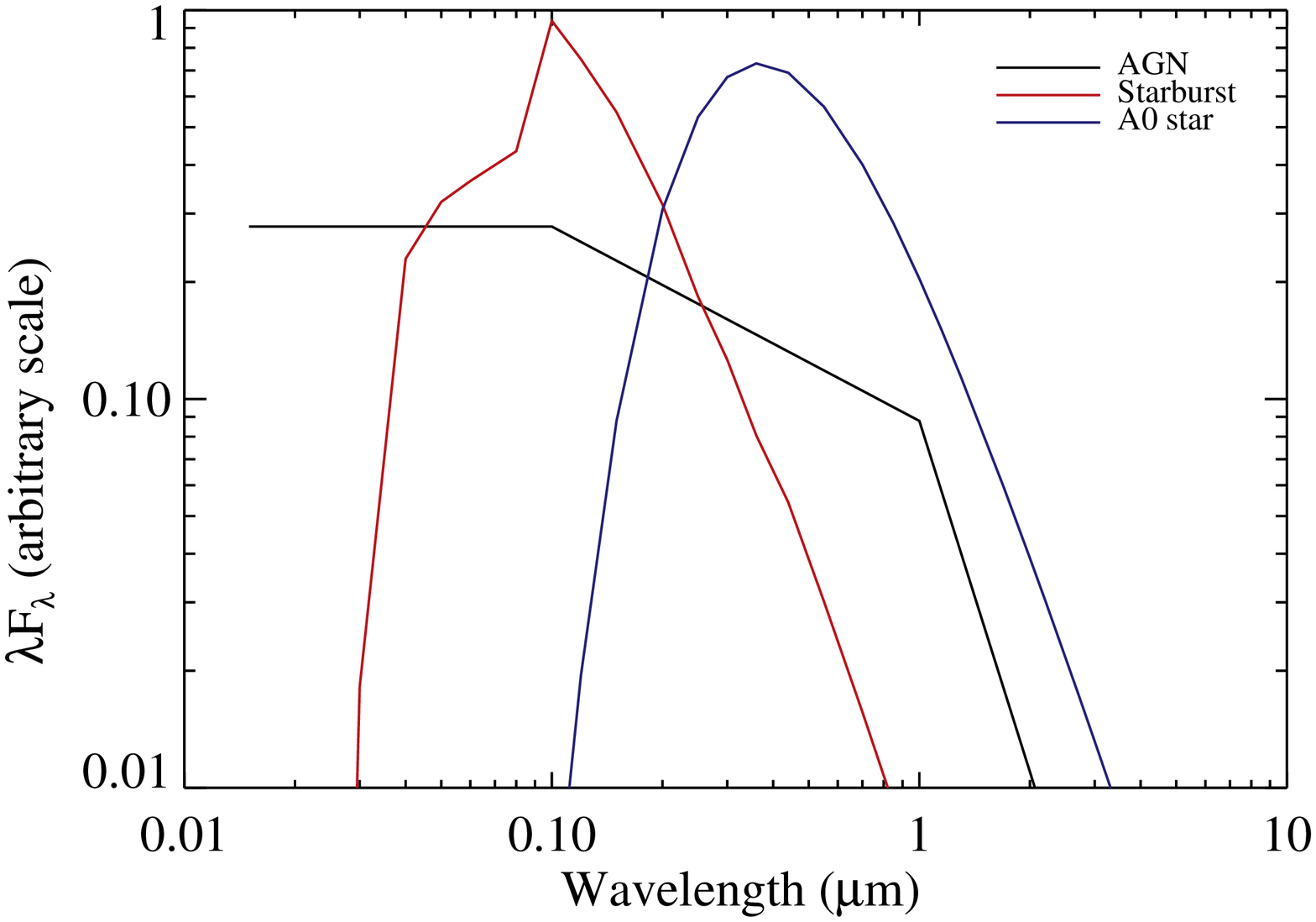}
\includegraphics[width=3in]{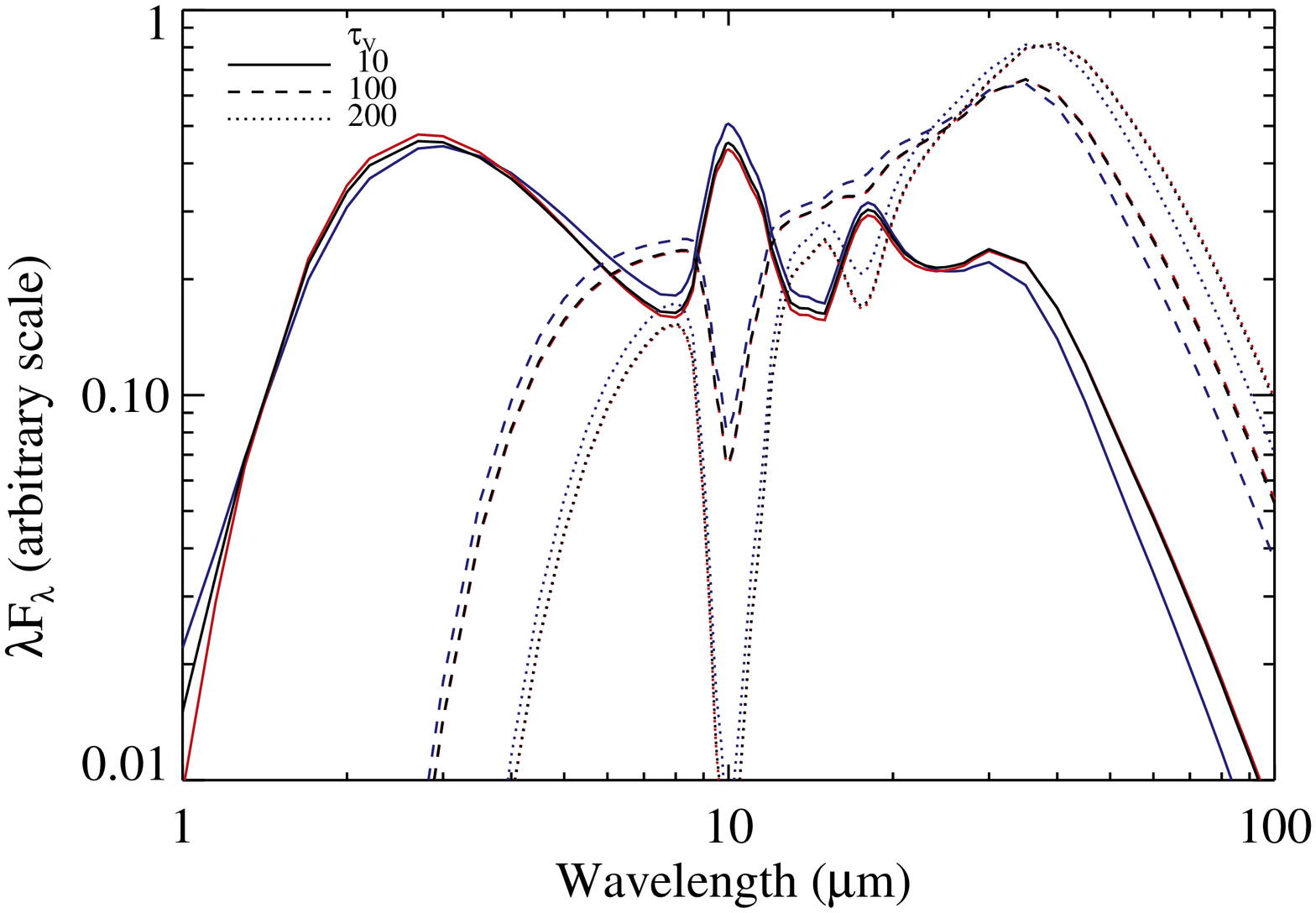}}
\caption{\label{fig:sedcomp}
Different heating spectra (\textit{left}) and their
resulting dust-reprocessed emission (\textit{right}); 
the emergent IR emission is
independent of the input spectrum. 
The three different input SEDs are normalized by their respective
bolometric fluxes.
Each is embedded in a smooth, geometrically thick shell of OHMc dust,
with density profile $\propto R^{-1}$, and shell thickness $Y = 500$.
The IR emission does not depend on the incident spectral shape
and requires only that the underlying spectrum provide some optical/UV or
shorter wavelength photons for heating.
Thus, for a given total optical depth, characterized by
$\tau_V$, the optical depth at 0.55\um, as marked,  
all three buried SEDs produce essentially the same emission.
The only evidence of the input SED 
appears below 1\um, 
for very low optical depths
($\tau_V < 1.0$), 
where scattering is significant.
The flux scale of both panels is the same, illustrating that
dust reprocessing dominates the resulting IR emission in all cases.
}
\end{figure}

\citet{Ive97} show that  for large optical depth, the 
resulting IR radiation 
is largely independent of the nature of the heating radiation. 
We consider three incident  spectral energy distributions (SEDs):
an AGN, a starburst, and a stellar blackbody
(Figure \ref{fig:sedcomp}, left panel).
The AGN spectrum is a broken 
power law described by \citet{Row95} and \citet{Nen02}.
We use Starburst99 \citep{Lei99, Vaz05} 
to calculate the SED of a
1-Myr-old instantaneous burst with solar metallicity
and a Salpeter initial mass function.
The right panel of Figure \ref{fig:sedcomp}
shows the results of the model calculation for smooth spherical
shells of OHMc dust with 
with profile $\propto R^{-1}$, shell thickness $Y = 500$, 
and various values of $\tau_V$. 
(We adopt these particular parameters because the resulting SEDs
vary dramatically as a function of optical depth.)
As is evident from the figure, there is no hint of the input
SED in the dust emission.  
The total optical depth of the dusty material
rather than the particular incident spectrum
determines the
emergent spectrum, even using the relatively cool A0 star, provided the
dust is close enough to be heated to 1500K.
In the subsequent calculations, we adopt the AGN incident spectrum; 
a luminous stellar population would yield equivalent results.

The same calculations underlie the 
models of continuous spherical shells we present here and
those of \citet{Ive97}.  Thus, their results that 
do not depend in detail on the dust chemistry, such as 
radial variations of temperature, are representative.
The comparison with their work demonstrates explicitly the
applicability of the computations and results we present here
to individual dust-embedded stars.  Although the disks and outflows of
young stellar objects can introduce asymmetry, 
the resulting SEDs remain similar 
(see Whitney et al. 2003 or Vinkovi{\'c} et al. 2003, for example). \nocite{Whi03}
\nocite{Vin03}
 
\section{Silicate Continuum and Feature Measurements\label{sec:cont}}

\subsection{Continuum Fitting\label{subsec:contfit}}
The silicate features appear characteristically different
in the nuclear spectra of various classes of galaxies.
Emission is typical of optically-identified quasars,
and deep absorption is typical of ULIRGs \citep{Hao07}.
To quantify these silicate properties, both in observations
and simulations, the strength
of the emission or absorption must be measured relative
to the underlying continuum.
A complication in the MIR is that dust is responsible for both the observed ``continuum'' and 
the spectral features; they do not have independent origins.

\begin{figure}[htb!]
\centerline{\includegraphics[width=6in]{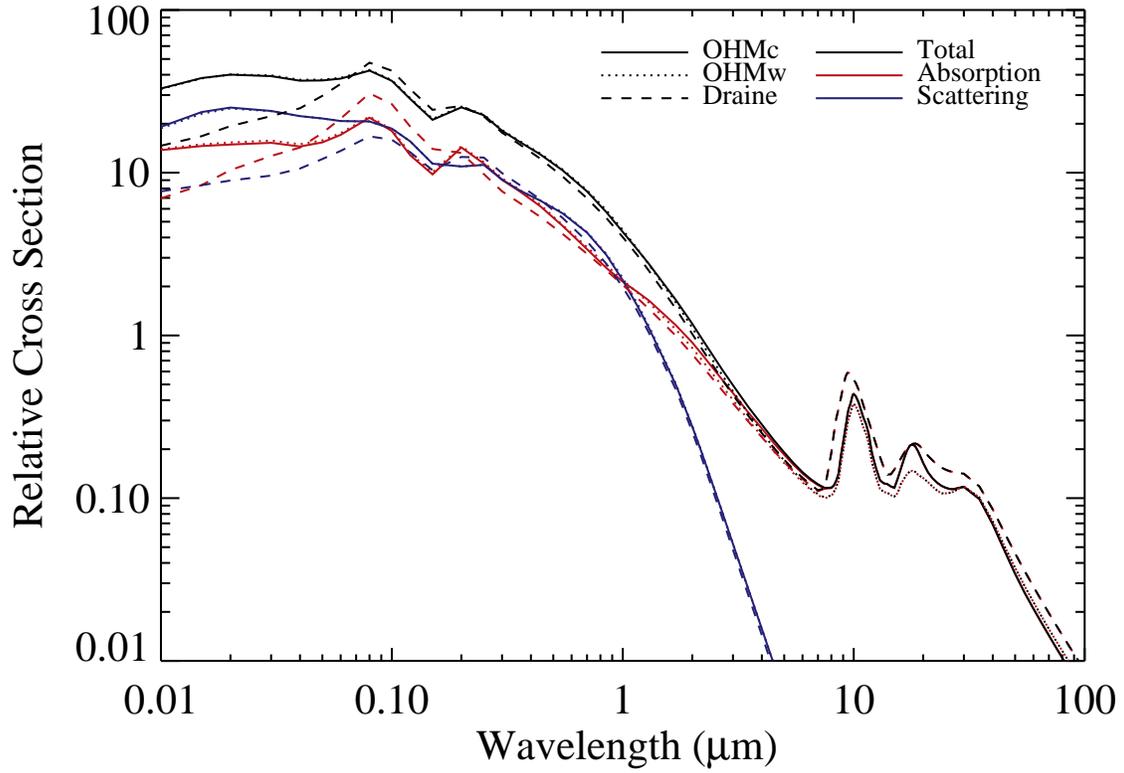}}
\caption{\label{fig:crossall}
Total, absorption, and scattering cross sections for the three
dust models we consider.  The total cross sections are
similar from 0.1 to 5\um.  However, 
the silicate features themselves and the ratio of the cross section 
at 10 and 18\um{} are
significantly different, which affects the resulting
spectra of dusty galaxies.}
\end{figure}

\begin{figure}[htb!]
\centerline{\includegraphics[width=3.5in]{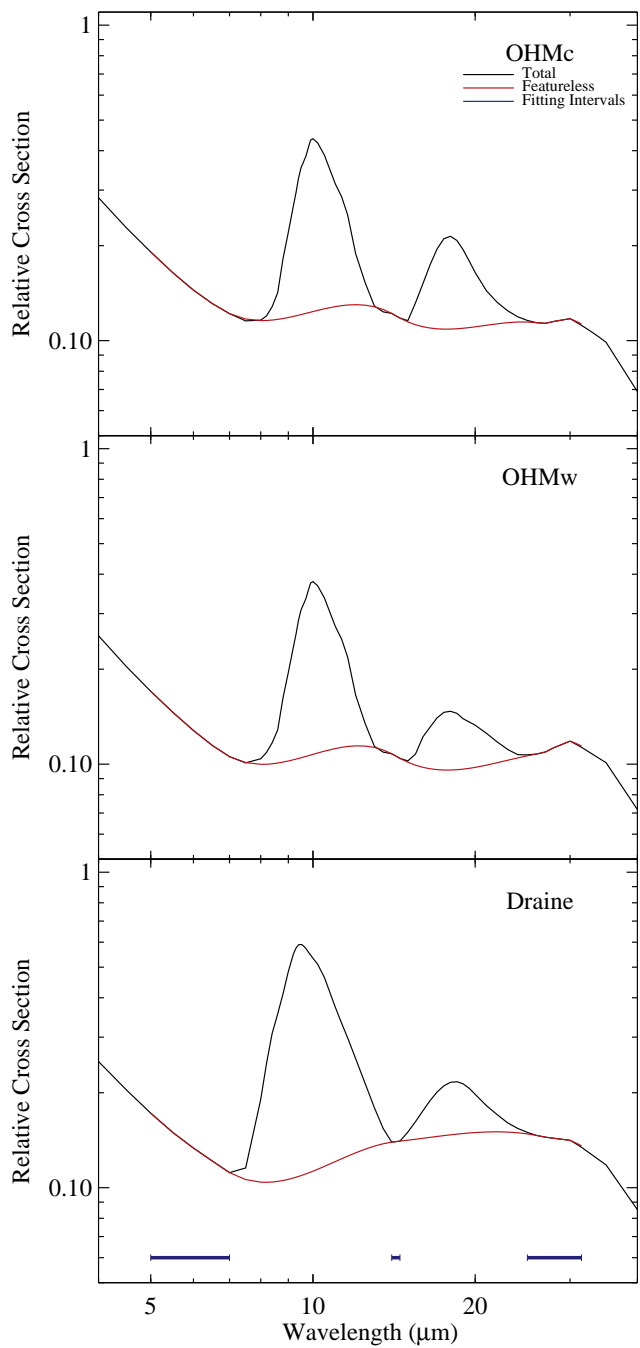}}
\caption{\label{fig:cross}
Total dust cross section as a function of wavelength
(\textit{black})  for three different dust chemistries. 
In all cases, two
prominent features stand out near 10\um{} and 18\um{}.
Each red curve is the spline interpolation of the total
cross section,
fit to only the intervals marked in blue. 
These fits
define the ``featureless dust,''
excluding the silicate features, from 5 to 31.5\um.
}
\end{figure}

Figure \ref{fig:crossall}  shows the total cross sections of the three dust models.
\citet{Oss92} tabulate cross sections
 only down to 0.4\um, 
and we  use DUSTY to extrapolate their results to shorter wavelengths employing 
the Mie theory\footnote{The published tabulation lists 63 wavelengths
between 0.4 and $10^4$\um.  We use 198 entries on a finer grid over the same 
spectral range kindly provided by T. Henning (1999, private communication).}.
Scattering is important 
at shorter wavelengths, but in the MIR, 
absorption dominates. 
The three  cross sections are  very
similar from 0.1 to 5\um{},
but they differ significantly in the
silicate features.  
For all dust chemistries, two features stand out
as local maxima 
in the cross section 
near 10 and 18\um.
Figure \ref{fig:cross}
shows the total (absorption and scattering)
cross section of this region for  the three dust models. 
The peaks  are not single-wavelength excitation but broad
 Si-O stretching and Si-O-Si bending modes, 
 which are
ultimately responsible for the emission and absorption
features observed in spectra.  
While all three models are empirical descriptions of
observed spectra, 
the profile and central
wavelength of each of these features and their
relative strength depend physically on the mineral composition of the dust.
These are evidently different in the three different dust models,
with the 18\um{} feature relatively stronger in the 
OHMc dust. 

In the familiar setting,
atomic or molecular lines are
superposed on a continuum, and different
 physical processes are responsible for the line 
and the continuum emission.
As a result, increasing the optical depth of lines 
leads to broadening and possible blending,
while the continuum remains fixed.
The properties of these lines, such as central wavelength
and spectral width, 
are well-defined, and they are independent of any continuum.
In contrast, in the case of grain emission,
the same  mechanism produces both the
features and the underlying continuum.
The features that are evident in spectra 
are a consequence of variations of the
dust cross section, but they are not independent entities.
In order to define measurable features, we
introduce synthetic ``featureless dust,''
which we obtain from  the
interpolation of the  smooth variation of 
the actual cross section.
We  define the features as the excess above this ``continuum.''
Figure \ref{fig:cross} shows these results for the
three dust chemistries.
In each case, the featureless dust, plotted in red, 
is a cubic spline interpolation of the total
cross section in logarithmic 
space over three intervals: 
 5--7, 14--14.5, and
25--31.5\um{}
(shown in blue). 
We still measure the continuum well at wavelengths longer than 31.5\um,
but we adopt this cutoff to develop a   
consistent method that we can apply to the 
spectra obtained with the IRS.

Figure \ref{fig:contfit} shows the result of
radiative transfer calculations for three representative
optical depths: $\tau_V  = 1$, 100, and 400.
At small optical depth, both features appear in emission,
and at large optical depth,  both are in absorption.
In the intermediate case, the 10\um{} feature is in
absorption while the 18\um{} is in emission.
In each panel, the 
red line shows the result of the
radiative transfer calculation for the exact same geometric
configuration, but using the featureless dust.
In order to quantify the strengths of the silicate features
in spectra, we employ the same technique to 
fit the underlying continuum:
cubic spline interpolation over the 
same three intervals shown in blue. 
The result is shown as the dashed blue line in each panel.
In each case, this line 
resembles the 
model result for the featureless dust. 
In other words, performing this interpolation procedure
over the cross section, to create ``featureless dust,''
or on the emergent spectrum, to measure the continuum,
produces the same result.
This outcome reflects a fundamental property of
radiative transfer: it proceeds wavelength-by-wavelength. 
Only an indirect coupling exists across wavelengths, 
through heating. 
This accounts for the
small differences between the red and dashed blue lines.
The emission or absorption at each wavelength is independent
of all others, 
so the features do not broaden  to occupy
a larger wavelength interval, 
 even at large optical depths. 
Instead, they remain confined to their original extent.

Including the intermediate-wavelength measurement is critical
to obtaining a good fit of the underlying continuum. 
The failure of continuum fitting that excludes the intermediate
interval (green dotted line) 
is evident mostly at moderate  and large
 visual optical depths.
In the former,
this alternate fitting technique almost entirely
misses the 18\um{} emission feature, and in the latter,
it greatly exaggerates the strength of both absorption
features.
We quantify the strengths of the features and give further
examples below (\S\ref{subsec:strength}).

\begin{figure}[htb!]
\centerline{\includegraphics[width=3.2in]{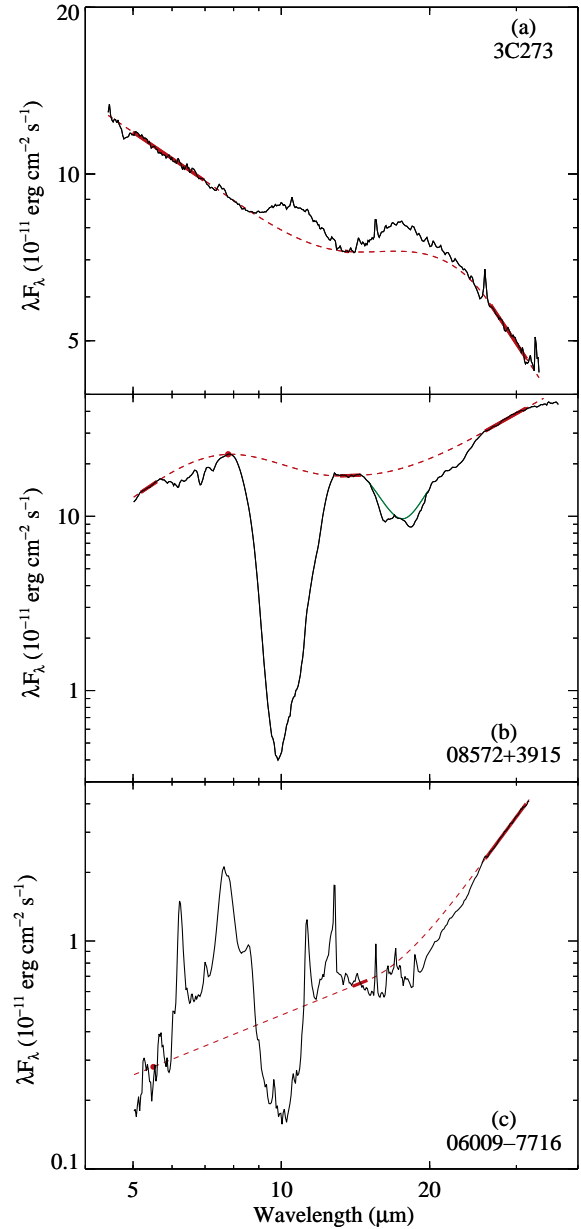}}
\caption{\label{fig:obs_methods}
Spitzer IRS observations  (\textit{black}) and continuum fits of
3C273, IRAS 08572+3915, and IRAS 06009-7716, which
serve as examples of
continuum-, absorption-, and PAH-dominated spectra (\textit{a}, \textit{b}, and \textit{c}, respectively).
Heavy lines mark the smoothed fitting intervals and dashed 
lines show the resulting continuum.
IRAS 08572+3915 shows a  significant crystalline silicate contribution around
18\um. The amorphous silicate component we identify 
in this case (see text)  is plotted in green.
(See Armus et al. 2008, in preparation, 
for spectra and continuum fits of the complete sample.)
}
\end{figure}

Application of this continuum-fitting method to 
spectra of galaxies is complicated by the presence of
emission lines, 
absorption by hydrocarbons and ices,
and emission of very small dust grains (PAHs).
Therefore, in order to measure the continuum in observations of galaxies, 
we must exclude these
emission and absorption features from the fitting intervals.
The technique we developed to overcome these problems was employed 
by
\citet{Spo07},  who outlined its essential points; here we
describe it in detail. 
We find three characteristic classes of spectra that require
modification of the fitting intervals.
Continuum-dominated spectra 
often exhibit high-ionization emission lines typical of
AGNs,
and they do not show any strong
PAH emission or ice absorption.
Water ice and aliphatic hydrocarbon absorption at shorter
MIR wavelengths (below 8\um) is strong
in the absorption-dominated spectra, 
and PAH emission remains weak.
The PAH emission is strong in the PAH-dominated spectra,
 with the equivalent width of the 6.2\um{} band, $EW_{6.2}$, 
exceeding 0.1\um, measured relative to the local continuum.
In all cases, to eliminate noise, 
power law fits of the observed flux within the selected
intervals, rather than the raw data, are used to fit the full continuum (Figure \ref{fig:obs_methods}).

Continuum-dominated spectra require only minor
modifications of the ideal fitting intervals.
To avoid the strong [\ion{Ne}{5}]$\lambda14.3\um$, the intermediate
fitting interval becomes 13.8--14.2\um.
When [\ion{O}{4}]$\lambda25.9\um$ is strong,
we increase the lower bound of the long-wavelength interval 
to 26.5\um.
We show the  continuum-dominated spectrum of 
3C273 and the resulting fitted continuum in Figure \ref{fig:obs_methods}a.

Strong absorption bands due to water ice (centered at
6\um), aliphatic hydrocarbons (at 6.90 and 7.25\um) and
gas phase C$_2$H$_2$ and HCN (at 13.7 and 14.0\um) 
characterize the absorption-dominated spectra.
(See \citealt{Spo05} for more detail.)
The short-wavelength interval is reduced to 
the 5.2--5.6\um{} region plus a connection point
at 7.8\um, and the intermediate- and long-wavelength
intervals become 
13.2--14.5 and 26.0--31.5\um, respectively. 
While PAH emission may be strong at 7.7\um, 
we do not find it to contaminate the 7.8\um{}
point in these spectra. 
Figure \ref{fig:obs_methods}b shows
the absorption-dominated spectrum of IRAS 08572+3915
and the resulting fitted continuum.

Direct measurement of the continuum is difficult 
when PAH emission is strong.
Most of the PAH emission is concentrated in two complexes
that extend from 6--9 and 10.5--14\um, with a 
minimum that coincides with the center of the 10\um{} silicate band.
Essentially none of the short-wavelength
flux within the Spitzer bandpass
of low-redshift galaxies  is true continuum in these cases.
We use a power law between the average of
the 5.3--5.7 and 14.0--15.0\um{} intervals with a 
spline fit extending through
the standard long-wavelength interval
to model the continuum.
We  consider this method for PAH-dominated spectra 
when $EW_{6.2} > 0.05\um$ and always employ it
when $EW_{6.2} > 0.1\um$.
We show IRAS 06009-7716, with strong PAH emission, and the subsequent
continuum fit in Figure \ref{fig:obs_methods}c as an example of
the method, although we exclude this and other ULIRGs that
exhibit strong PAH emission from the sample below.

\subsection{Feature Strengths\label{subsec:strength}}

We define the strength of a feature as the natural logarithm of the 
ratio of the total emission to the underlying continuum:
\eq{ S = \ln \frac{f_{obs}(\lambda_m)}{f_{cont}(\lambda_m)},  \label{eqn:sdef} }
 evaluated at $\lambda_m$, where the strength is an extremum. 
Both  OHMc and OHMw dust cross sections have local maxima at 10.0 and 18.0\um,
while the Draine dust  has local maxima at 9.5 and 18.5\um.
We use 
the extremum of $S$ near these wavelengths
to measure the 
10 and 18\um{} strengths,
$S_{10}$ and $S_{18}$, respectively.
  Positive values of 
$S$ indicate emission above the continuum, and negative
values of $S$ indicate absorption.
When the silicate feature is in absorption, the
definition of $S$ 
corresponds
to {\em apparent} optical depth; 
\textit{i.e.}, $S$ is negative, so
$f_{obs}= f_{cont}e^{-{\tau}_{app}}$, with $\tau_{app} = -S$. 
However, we emphasize that 
this calculation {\em does not} yield the actual optical depth along the
line of sight.
For example, in Figure \ref{fig:contfit}c,
at 10\um\ 
the dust optical depth is 
$\tau_{sil, 10} = 17$, yet the apparent optical depth 
is only 1.6.
Furthermore, the apparent and actual optical depths are not
related to each other in any direct way because the strength depends on
the geometric distribution and is not a single-valued function 
at extremely large optical depth
\citep{Lev07}.

The wavelength at which $S_{10}$  or $S_{18}$
is evaluated, $\lambda_{m}$, is not fixed but is
rather the wavelength of the local extreme strength. 
In the simulations, we find $\lambda_m$ of the shorter-wavelength
feature typically near the wavelength of the peak
cross section, although ranging from 9.2--10.0, 9.3--10.2, and
9.0--9.8\um, respectively, when the OHMc, OHMw, or Draine dust is used.
The longer-wavelength feature 
typically  peaks 
close to 18.0\um, ranging from 16.0 to 19.5\um{}
in all simulations.
The former is often described as the ``9.7\um''
feature, with the
measured center 
typically varying from 9.7 to 10.0\um{} in absorption
and often exceeding 10\um{} in emission \citep[\textit{e.g.},][]{Hao05}. 
In observations, we evaluate $S_{10}$ at the wavelength of the
local extremum relative
to the continuum between 8 and 13\um, $\lambda_{10}$,
and we measure $S_{18}$ at $\lambda_{18}$, the  wavelength of the local extremum
between 15 and 22\um.
Crystalline silicates
can produce strong features at 17 and 20\um, especially in
absorption-dominated spectra.
In these cases, we identify the amorphous contribution in
a spline fit to the data in regions free of
crystalline absorption, and we evaluate 
$S_{18}$  in the amorphous component alone \citep{Spo06}. 
Figure \ref{fig:obs_methods}b 
shows the spectrum of IRAS 08572+3915
 (in black), an example  of a strong crystalline component,
 and the green line shows the 
amorphous silicate contribution we identify.

We apply these methods to the ULIRGs observed with the IRS
that \citet{Spo07} and \citet{Ima07} have presented. 
We restrict consideration to those galaxies that have
$EW_{6.2} < 0.1$.  When
$EW_{6.2}$ is larger, continuum measurements are less certain.
This selection criterion excludes a number of familiar
ULIRGs, such as Arp 220 and Mrk 273.
We caution that this sample selection
does not allow us to draw  conclusions about the
nature of ULIRGs' energy sources generally,
for we reject the objects that exhibit the 
strongest direct evidence for dominant star formation. 
We also eliminate all spectra that do not have complete coverage of
both the 10\um{} and 18\um{} features.
The resulting sample (of 46) 
and their silicate strengths are listed
in Table \ref{tab:obsv} by their IRAS identification or other common name.
Within the sample, where the continuum fitting is robust,
the order-to-order flux uncertainty of the spectra themselves
 dominate the errors in the 
strength measurements, 
which are typically less than 10\%.
Uncertainties of
the  continuum fitting are 
greater  
in the sources with strong PAH emission, which we exclude from this study,
resulting in larger errors in the strength measurements.

In all cases, 
including the intermediate-wavelength continuum data
is critical to obtaining reliable strength measurements.
In quasar spectra in which both features are detected
in emission, neglecting the intermediate-wavelength data
in the fit suppresses the long-wavelength
continuum, resulting in larger 18\um\ strengths.
In some cases, \textit{e.g.}, in the fits of \citet{Hao05}, 
$S_{18}$ 
even becomes much greater than $S_{10}$.
For example, in
3C273, they find
$S_{10} = 0.1$ and $S_{18}= 0.3$.
Including the intermediate-wavelength
continuum measurement reduces this ratio;
we  find both 
$S_{10}$ and $S_{18}= 0.1$.
\section{Diagnostics of Dust Geometry and Chemistry}

\subsection{Silicate Feature Strength\label{subsec:oneline}}

\begin{figure}[htb!]
\centerline{\includegraphics[width=3.in]{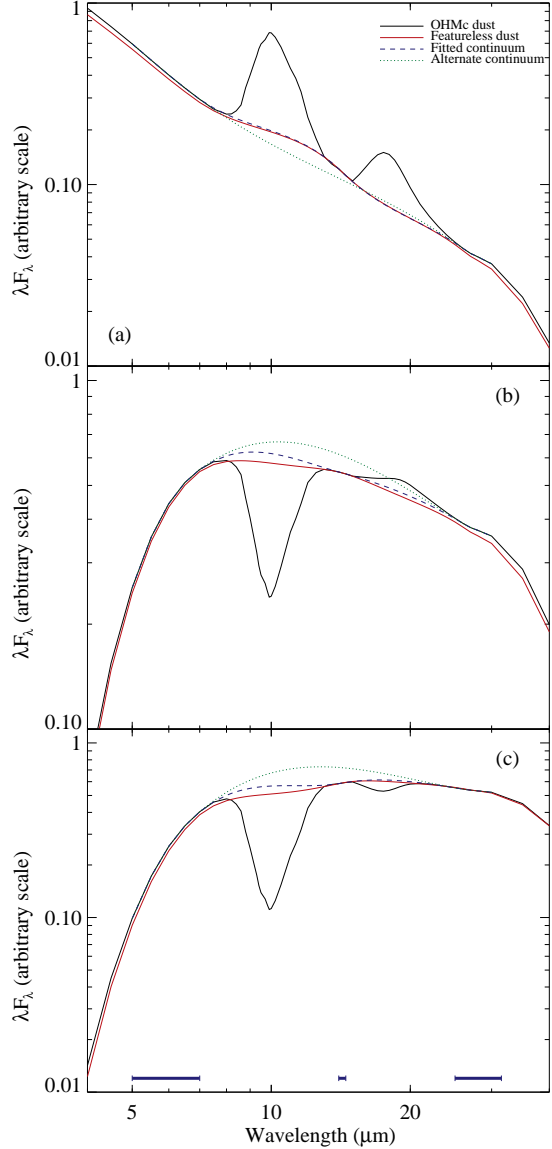}}
\caption{\label{fig:contfit} 
Numerical calculation of the dust-reprocessed
emission (\textit{solid black line}) 
due to an AGN located in a spherically-symmetric
continuous distribution of OHMc 
dust, with radial density
profile $\propto R^{-2}$, and shell thickness $Y = 600$.
Results for $\tau_V = 1$, 200, and 400 are plotted (\textit{top to bottom}). 
Adopting the ``featureless dust'' cross section (the red line of
Figure \ref{fig:cross}) while retaining
the same dust geometry and heating spectrum
produces the spectrum plotted in red.
Fitting a spline over the three selected intervals (marked in blue 
at the bottom of the plot)
yields the continuum plotted as a {blue dashed line}. 
The good agreement between the ``featureless dust'' spectrum
and this fitted continuum demonstrates that this 
method, 
including an intermediate-wavelength measurement between the
10 and 18\um{} features, describes
the underlying continuum well.
In contrast, fitting only the outer regions (below 7 and above 25\um)
results in a poor continuum model (\textit{green dotted line}), 
and consequently,
the strength of the silicate features cannot be measured accurately.}
\end{figure}

Dust reprocessing  of intrinsic radiation is directly
responsible for the entire observed MIR spectra of ULIRGs.
Thus, the spectra are sensitive to dust properties
in detail, and both the total optical depth of the dusty
material and its geometric distribution determine
the behavior of the silicate features. 
Consider 
 a dusty region having  optical depth, 
$\tau_\lambda$,
and consider first the case of
constant temperature. 
The emission from such a region
is $I_\lambda = 
B_\lambda(1 - e^{-\tau_\lambda})$, where $B_\lambda$ is the Planck function
of the temperature. 
When $\tau_{sil}\ll 1$, $I_\lambda \simeq B_\lambda \tau_\lambda$.
Since $\tau_\lambda$  is proportional to the dust cross section,
 in this case 
the emergent spectrum
 has the same shape as the profiles
shown in Figures \ref{fig:crossall} and \ref{fig:cross}.
Therefore, the feature is in emission ($S> 0$)
and it has the same strength as 
the plotted cross sections, 
 independent of optical depth. (For an example, see Figure \ref{fig:contfit}a.) 
When $\tau_{sil}$ increases and approaches unity, 
self-absorption sets in, and 
the feature strength decreases.
As $\tau_\lambda$ increases further and exceeds unity, 
$I_\lambda$ becomes equal to $B_\lambda$:
at constant temperature, self-absorption and emission exactly balance each other,
producing the thermodynamic limit of the Planck function.
Therefore, for all $\tau_{sil}>1$, a {\em single}-temperature region can never produce
a feature, either in emission or absorption.
The emergence of an absorption feature (Figures \ref{fig:contfit}b and c)
reflects the temperature stratification in actual dusty material: 
as the radiation propagates from  hot regions
toward the observer, it passes through  cooler regions where
it  suffers absorption that is not balanced by
the emission from these cooler regions.
Thus, the temperature structure primarily determines the
strength of the absorption feature at large optical depths. 
The  10 and 18\um{} features display a different behavior,
with the former switching from emission to absorption at
a smaller total optical depth, because
$\tau_{sil,10} > \tau_{sil,18}$ (Figure \ref{fig:contfit}b).

Either radiative transfer effects or  geometric dilution can produce
temperature gradients.
In the first case, 
the surface layer absorbs the heating radiation,
and the propagation to subsequent layers degrades the
photons to longer wavelengths.
Radiative transfer can produce only a limited temperature difference,
which is similar in all geometries.
In contrast, 
the spatial dilution of radiation is unique to the
geometrically thick dust shell around a central heating source.
Indeed, \citet{Lev07} 
show that
deep silicate absorption requires dust distributions that are 
not only optically but also geometrically thick.
However, the $p=2$ 
models never show large silicate strength; 
although the shell may be formally geometrically thick,
the
dust is concentrated, with 80\% located within $R\le 5R_d$.
The slab and cloudy models are confined to the
low silicate strength regime \citep{Nen02,Nen07}.

The cross section at either the 10 or 18\um{} silicate feature
depends on the mineral composition of the dust.  
(See \citealt{Mol03} for a review.) 
Various
silicates have different maximal wavelengths for 
absorption.  Combinations of multiple silicates alter
the breadth of the features immediately in the cross section,
with consequential effects in the observed spectra.
Altering the 
size distribution of grains affects primarily their scattering
properties.
Thus, the grain sizes do not strongly 
influence the feature strengths in the MIR, where absorption is most important.

\begin{figure}[htb!]
\centerline{\includegraphics[width=3in]{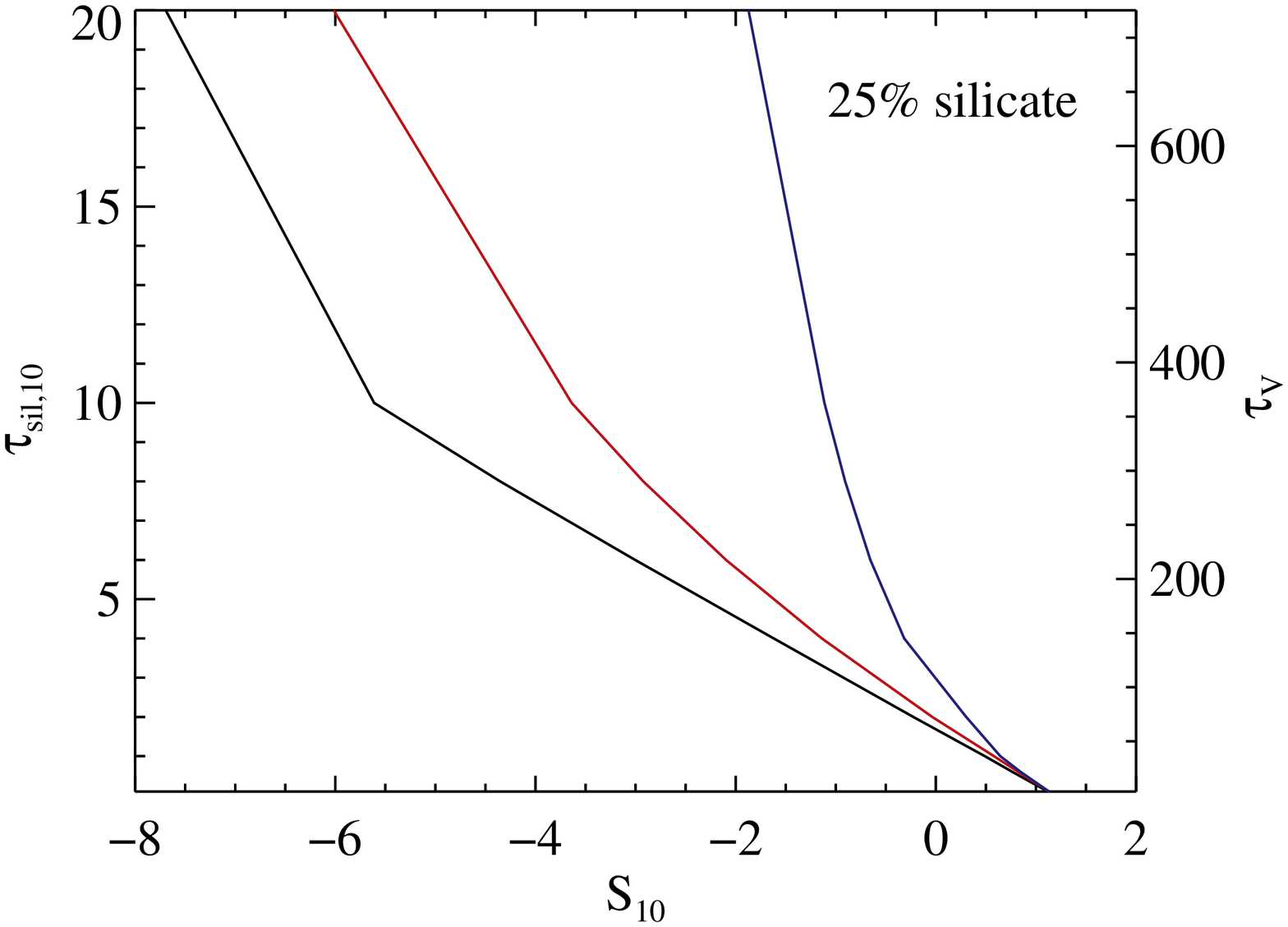}
\includegraphics[width=3in]{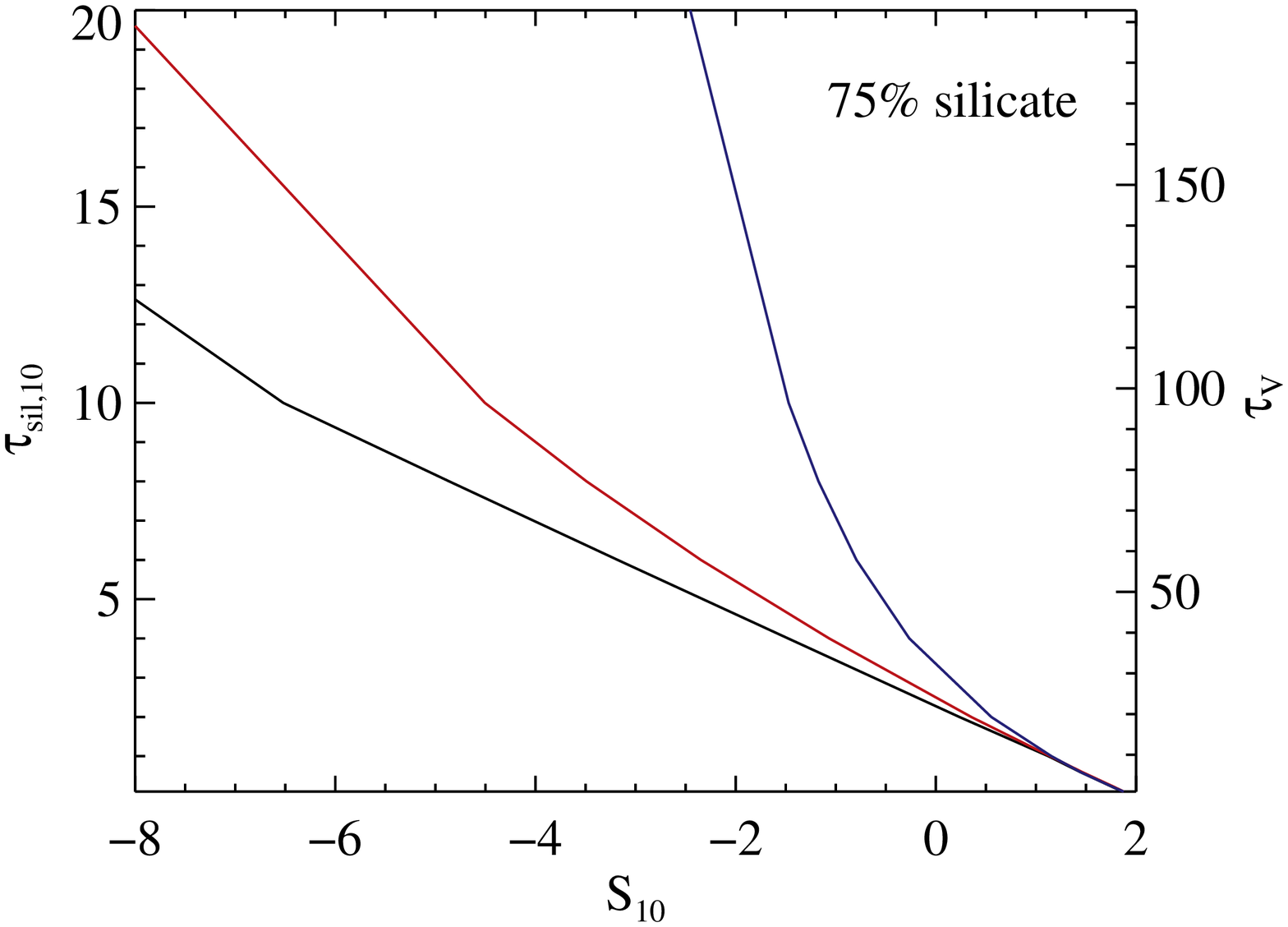}
}
\caption{\label{fig:abun}
Relationship between 
feature strength, $S_{10}$, and
$\tau_{sil,10}$, the 
 optical depth at the peak of the 10\um{} silicate feature, 
for 
silicate abundance of 25\% (\textit{left}) and 75\% (\textit{right}).
The corresponding optical depth at $0.55\um$, $\tau_V$, is marked 
along the right axis.
Results are calculated for $Y=200$, with $p = 0$, 1, and 2
plotted in black, red, and blue, respectively.
For the two abundances, the
 results are similar as a function of
$\tau_{sil,10}$ because  $S_{10}$ fundamentally
depends on the absolute amount of silicate present.
For example, when $\tau_{sil, 10} = 4$ (which corresponds to $\tau_V = 150$ 
in the low-abundance model, 
and
$\tau_V = 40$ 
in the high-abundance model) 
the 
 feature strength is the same in both panels for each density distribution.
For $p=2,$ both panels display  virtually identical results at 
all large optical depths. 
The significance of heating effects increases as the
density distribution becomes shallower, introducing slight
differences between the two abundance calculations. 
Large silicate strengths do not indicate high silicate abundances
but are instead 
very sensitive to the geometric distribution of dusty material.
}
\end{figure}

For a given mineral composition, changing the silicate abundance
relative to carbon simply results in changing the scale
of strength  relative to the optical
depth at 0.55\um.  Because the silicate produces both the MIR continuum
and features, the absolute amount of silicate 
sets the feature strength.
Thus, only the scaling of $S_{10}$  with $\tau_V$
is altered with a change in abundance.
Figure \ref{fig:abun} shows the scaling relations for
two extreme abundance variations and demonstrates 
that the 
optical depth of silicate, not the silicate abundance,
governs the feature strength. 
Increasing the silicate abundance does not increase the absolute feature 
strength, so strong features, whether in emission or absorption,
are not a signature of high silicate abundance.

\subsection{Feature-Feature Diagram \label{subsec:ratio}}

The different optical depths around  10 and 18\um{} 
result in differing behavior of the two silicate features,
and their appearance  depends on the geometric dust configuration.
In order to  examine these systematic differences 
we introduce the  ``feature-feature diagram,'' in which we 
plot $S_{18}\ vs.\ S_{10}$. 
For a given dust chemistry, irrespective of the geometry, 
all optically thin configurations
are located at the same point in the diagram, 
corresponding to the strengths of the 
two features in the
dust absorption coefficient.
For each dust model, the coordinates of this optically thin point are listed 
in Table \ref{tab:dust},
where the notation $S^c$ refers to the silicate strength measured in
the cross section.
Increasing  the optical depth 
of a given dust geometry
produces tracks that 
move away from this point toward deeper absorption.
Figures \ref{fig:s18v10ohmc} through \ref{fig:s18v10d}
show the feature-feature diagrams
for the three different dust chemistries.
In each diagram, we plot
the tracks for the basic geometries of slab and 
clumpy and smooth spherical shells.
As described above, only the smooth  shells
ever exhibit large feature strengths.
Remarkably, for each dust composition, all smooth shells exhibit 
nearly the same behavior, independent of the
density distribution:
tracks have nearly the same slope,
set by the optically thin feature strengths.
In other words, the optical properties of the dust
set both the initial point and the slope of the tracks.

The similarity of the smooth shell tracks shows the
dominant effect of chemistry, and 
the scatter of these trends is a consequence of 
the differing temperatures that arise at the $\tau=1$
surface where each observable feature is produced.
In general, in the geometrically thicker shells, where the dust is spread over a larger
volume, the tracks are steeper.  
The same optical
depth produces greater strength because 
of the larger temperature  gradient (\S \ref{subsec:oneline}).
This effect is most evident
for $p=0$ and smaller for $p=1$, while
 the $p=2$ tracks are independent of $Y$.
In this last case, 
 the dusty material is concentrated in the inner regions, and the 
observed $\tau=1$ surface is at 
constant temperature in all shells.
These tracks also stand out because they
never produce large feature strengths, ending closer to the
optically thin point at $\tau_V = 400$, the maximum plotted. 

The clumpy models yield results similar to those of the slab
on the feature-feature diagram.
None of the clumpy configurations exhibit the extremely deep
absorption of the continuous thick shells,
and for a given value of $S_{18}$, the 10\um{} absorption
is weaker compared with the smooth geometry.
For clarity, we plot  only the results for
$N_0=1$, 3, and 5, fixing $p=0$ and $Y=30$. 
As with all other distributions, these tracks originate
at the optically thin point.  However, as the
optical depth of a single cloud, $\tau_{V,C}$, 
 increases, these
tracks diverge away from those of the smooth shell models.
The plotted clumpy results begin at
$\tau_{V,C} =10$, so 
the total optical depth on each track's upper right
is $10 N_0$.
These and all subsequent clumpy results
 lie in a region inaccessible to the smooth shell tracks.
Along each track, $\tau_{V,C}$ increases to a maximum of 80, 
for total optical depth $80 N_0$.
Individual tracks extend farther to the lower left
as  $N_0$ increases, and they become more similar
to the continuous shell results when $N_0$ is very large,
\textit{i.e.,} as the 
geometric distributions become physically indistinguishable.

As optical depth increases, 
both $S_{10}$ and $S_{18}$  initially follow the trends of
the slab, moving away from the optically thin limit. 
When the dust becomes optically thick in the
silicates, however, the strength
becomes less negative.  
In contrast to the slab, the clumpy 
obscuration offers the possibility of observing some bright,
directly-illuminated cloud
faces.  The silicate emission from these bright faces fills 
in the absorption trough, reducing the depth of the 
absorption feature in the spectrum \citep{Nen02}.
These optical depth effects first become
relevant to the 10\um{} feature, producing the 
curvature evident in  the resulting tracks.
Although both the
clumpy and smooth configurations can result in
weak $S_{10}$, the combination of both silicate features
together breaks this degeneracy and can be used to discriminate between these
two fundamentally different geometries.

\section{Observations and Dust Chemistry\label{sec:chem}}

\subsection{ULIRGs and Local Sources\label{subsec:data}}
\begin{figure}[htb!]
\centerline{\includegraphics[width=6in]{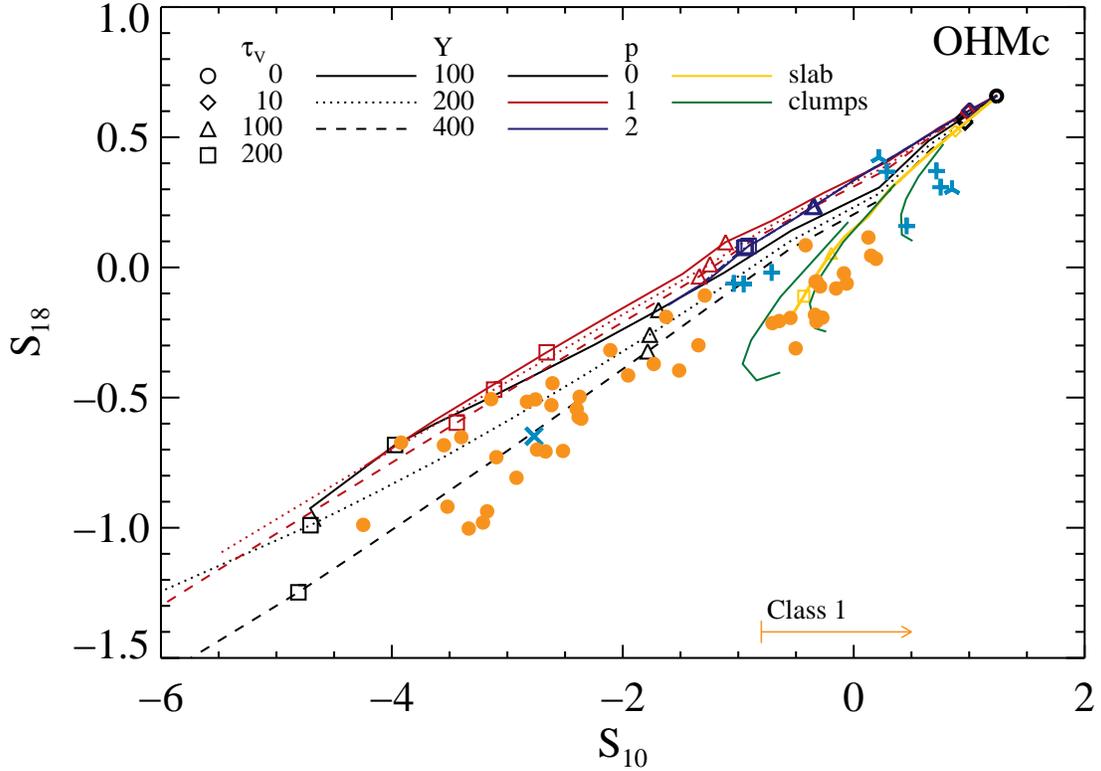}}
\caption{\label{fig:s18v10ohmc}
Strength of the 18\um{} silicate feature \textit{vs.}
strength of the 10\um{} silicate feature computed in
numerical simulations using OHMc dust (\textit{lines}) 
and measured in ULIRGs (\textit{filled symbols}).
Computations of different geometries are plotted in different
colors, and open symbols along these lines mark 
total optical depth.  In the smooth spherical shell models, line
style indicates shell thickness, $Y$, and color indicates
radial power of the density distribution, $p$.
For the clumpy models, we plot 
only the $p=0$, $Y=30$  results.
Each  track corresponds to a different value of 
$N_0$  (with  1, 3, and 5 from the upper right to lower left, respectively).
Along each track, $\tau_V$ for a single cloud increases 
away from the optically thin point to a maximum of 80. 
For comparison, a number of  lines of sight 
in the Galaxy and Magellanic Clouds
are also plotted (\textit{cyan skeletal symbols}).
Galaxies having $S_{10}>-0.8$ are identified as ``Class 1''
in the  Spoon et al. (2007) scheme. 
The dust composition sets the values of
$S^c_{18}$ and $S^c_{10}$ listed in Table \ref{tab:dust}, which are
the origin of all tracks (the limit of $\tau_V \to 0$).
The smooth spherical shell models generally exhibit a linear
trend in $S_{18} \ vs.\ S_{10}$, with greater strength
at larger optical depth.  
Variations of the thermal gradient as a function of
dust geometry are responsible for the scatter
in this relationship.
This dust composition, having relatively high ratio of
absorption cross section at 18 to 10 \um, 
is consistent with the observations, even
when the strengths are large.
}
\end{figure}

\begin{figure}[htb!]
\centerline{\includegraphics[width=6in]{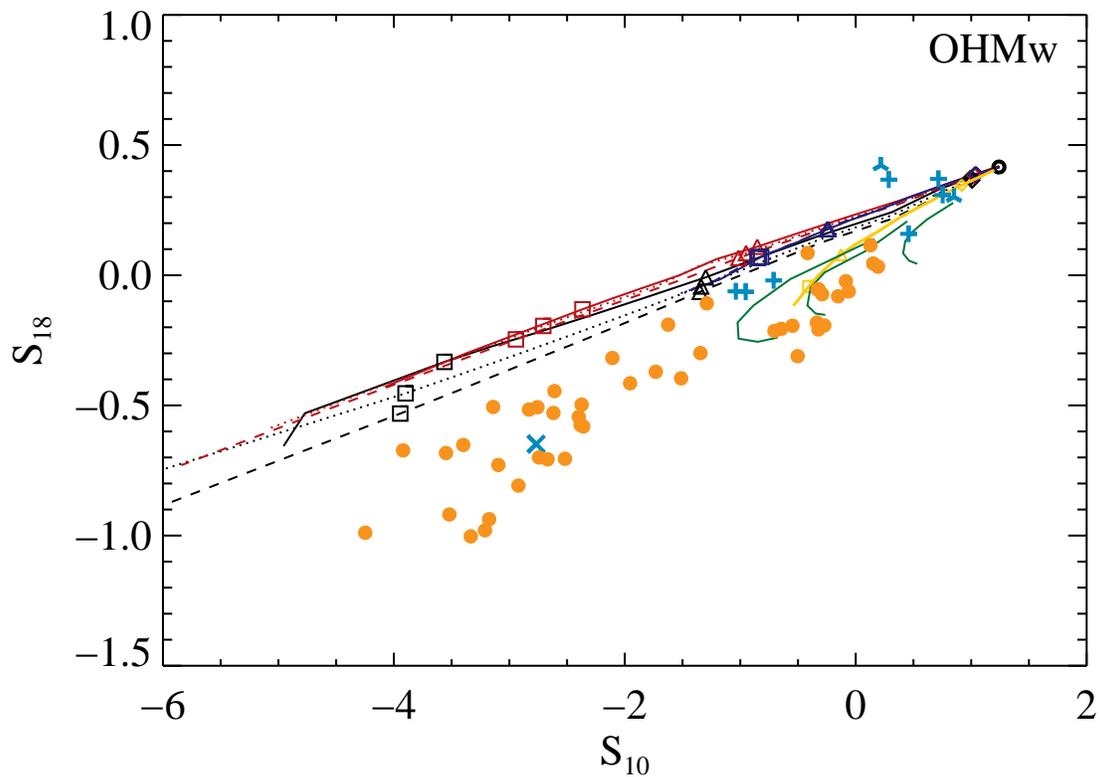}}
\caption{\label{fig:s18v10ohmw}
Strength of the 18\um{} silicate feature \textit{vs.} 
strength of the 10\um{} silicate feature computed in
numerical simulations using OHMw dust (\textit{lines}).
Other lines and symbols as in Figure \ref{fig:s18v10ohmc}.
Simulations using this 
dust chemistry cannot account for the observed deep absorption of ULIRGs
and also cannot accomodate
the relatively strong 18\um\ emission of local sources, 
which fall above the tracks,
despite the large range of geometric
configurations calculated.
}
\end{figure}

\begin{figure}[htb!]
\centerline{\includegraphics[width=6in]{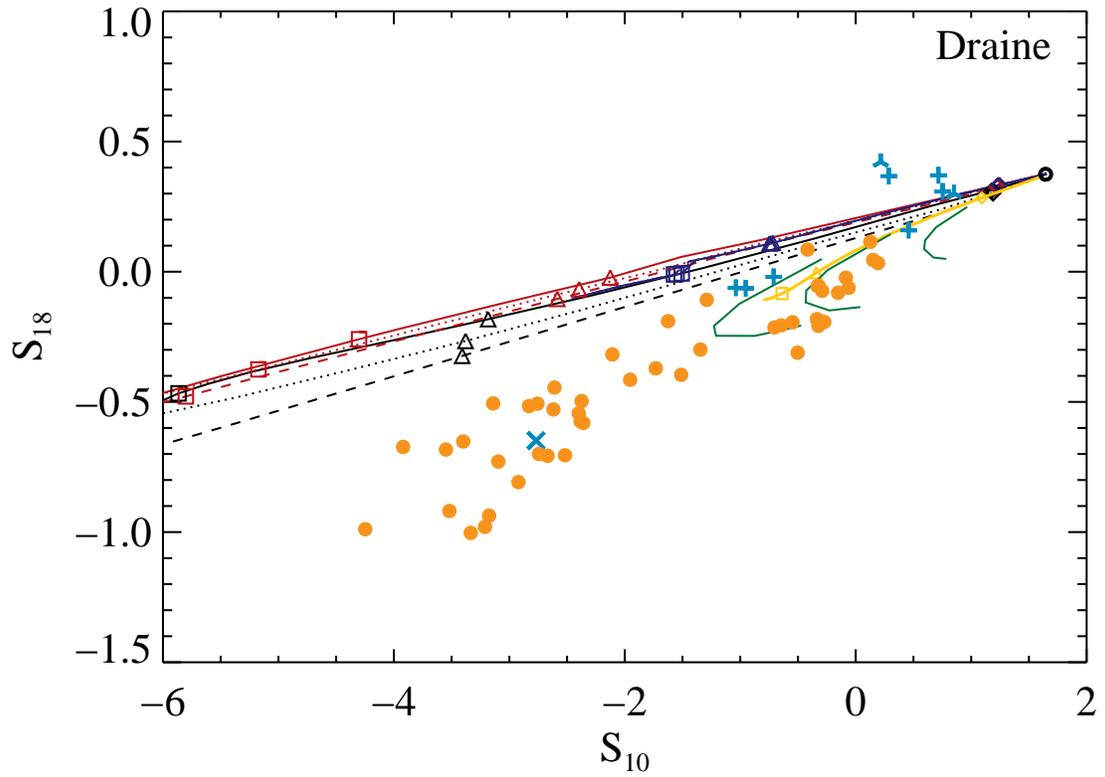}}
\caption{\label{fig:s18v10d}
Strength of the 18\um{} silicate feature \textit{vs.} 
strength of the 10\um{} silicate feature computed in
numerical simulations using Draine dust (\textit{lines}).
Other lines and symbols as in Figure \ref{fig:s18v10ohmc}.
This dust chemistry shares the
same problems of the OHMw dust (Figure \ref{fig:s18v10ohmw}).
}
\end{figure}

The ULIRGs we observed exhibit the same general trend of 
increasing $S_{18}$ with $S_{10}$, plotted as filled symbols on
Figures \ref{fig:s18v10ohmc} through \ref{fig:s18v10d}. 
In detail, however, the
measured ratios deviate from the simulations that use either the
OHMw or Draine dust, particularly at great strengths where
large optical depths are required.  
Neither of these models can reproduce the observed
 $S_{18}/S_{10}$ ratio when $S_{10} < -1$.
Instead, the data are consistent
with the OHMc dust model, in which both
$\tau_{sil,18}/\tau_{sil,10}$ and
$S^c_{18}/S^c_{10}$
are relatively higher (Table \ref{tab:dust}). 

The comparison with a number of objects in the Galaxy and Magellanic Clouds demonstrates
that the ULIRGs do not have unusual dust properties.  Indeed, the similarity
of all sources shows that the mineralogy that the
OHMc dust represents is appropriate in both Galactic and extragalactic contexts.
We include MIR observations of several planetary nebulae and giant \ion{H}{2} regions 
(\citealt{Pot05}; \citealt{Leb07}; 
 Bernard-Salas et al., in preparation; Lebouteiller et al., in preparation;
Whelan et al., in preparation),
and the Wolf-Rayet star GCS 4 
\citep{Gib04} on 
Figures \ref{fig:s18v10ohmc} through \ref{fig:s18v10d} 
(plotted as $+$, \rotatebox[]{180}{\textsf{Y}}, and $\times$, respectively).  
The last of these offers the advantages of a lack of circumstellar dust
and a well-behaved
 intrinsic continuum, which can be modeled accurately to yield the
interstellar absorption \citep*{Kem04}.

The local sources provide  further 
evidence in favor of the OHMc dust. 
Several of the low optical depth
local sources lie {\em above} 
all the Draine and OHMw tracks, 
and no geometry can accommodate such a location.
The  clumpy results, which are not very sensitive to chemistry,
cover the region {\em below} the smooth shell tracks, 
but 
the region above  these tracks is inaccessible
in any  given dust chemistry.
Overall, only the OHMc tracks can successfully 
describe {\em all} the  low optical depth sources
with  smooth or clumpy shells.

Other studies also provide evidence in favor of the OHMc dust.
\citet{Rou06} use DUSTY to fit the IR SED of the 
nascent starburst galaxy NGC 1377, and 
their best-fitting model employs this dust chemistry.
The adoption of the OHMc optical properties is critical to  
their successful description of the Spitzer IRS spectrum. 
The dust is actually more important
than the multiple free parameters they use to describe the geometry.
Furthermore, 
the OHMc dust better fits 
the central wavelengths of the observed silicate features,
with the observed
10\um{} absorption peak  arising at a longer wavelength
than the Draine dust would produce.

Other dust models to consider are those of \citet{Chi06}, who
also prefer the longer-wavelength 10\um\ silicate peak.
These authors find  relatively strong  interstellar absorption at 18\um{} 
in observations of several Galactic Wolf-Rayet stars,
with $\tau_{sil,18}/\tau_{sil,10}=0.55$ toward the Galactic
Center,  which is
comparable to the corresponding OHMc ratio,
and 0.70 in the local ISM. 
Independently, measurements of molecular hydrogen line ratios along lines
of sight near the Galactic Center also favor similar high 
values for this  ratio \citep{Sim07}.
In contrast, 
the \citet{Chi06} extinction curves
have
 rather low strength ratios ($S^c_{18}/S^c_{10} = 0.29$ and 0.19,
for the Galactic Center and local ISM, respectively).  The
former is 
intermediate between  the Draine and OHMw values, while the latter
is lower than all others we consider. 
We did not employ either of the \citet{Chi06} dust models in
the computational models, which require 
complete tabulations of the scattering and absorption 
cross sections. 

\subsection{Artificial Dust Chemistry\label{subsec:artdust}}

We consider whether modifying the 
 tabulated optical properties of 
dust in a simple way
can bring all the resulting smooth shell 
tracks into agreement with the large strength observations.
We artificially scale the 
cross sections of the two silicate features by the same factor  
relative to
the featureless continuum. 
This procedure is equivalent to scaling the excess over the red curves
plotted in Figure \ref{fig:cross}.
The ratio $S^c_{18}/S^c_{10}$ remains constant,
although both  $S^c_{18}$ and $S^c_{10}$ individually change.
Because we do not alter the underlying continuum,
this transformation {\em does} change
the ratio
$\tau_{sil,18}/\tau_{sil,10}$.  

\begin{figure}[htb!]
\centerline{\includegraphics[width=6in]{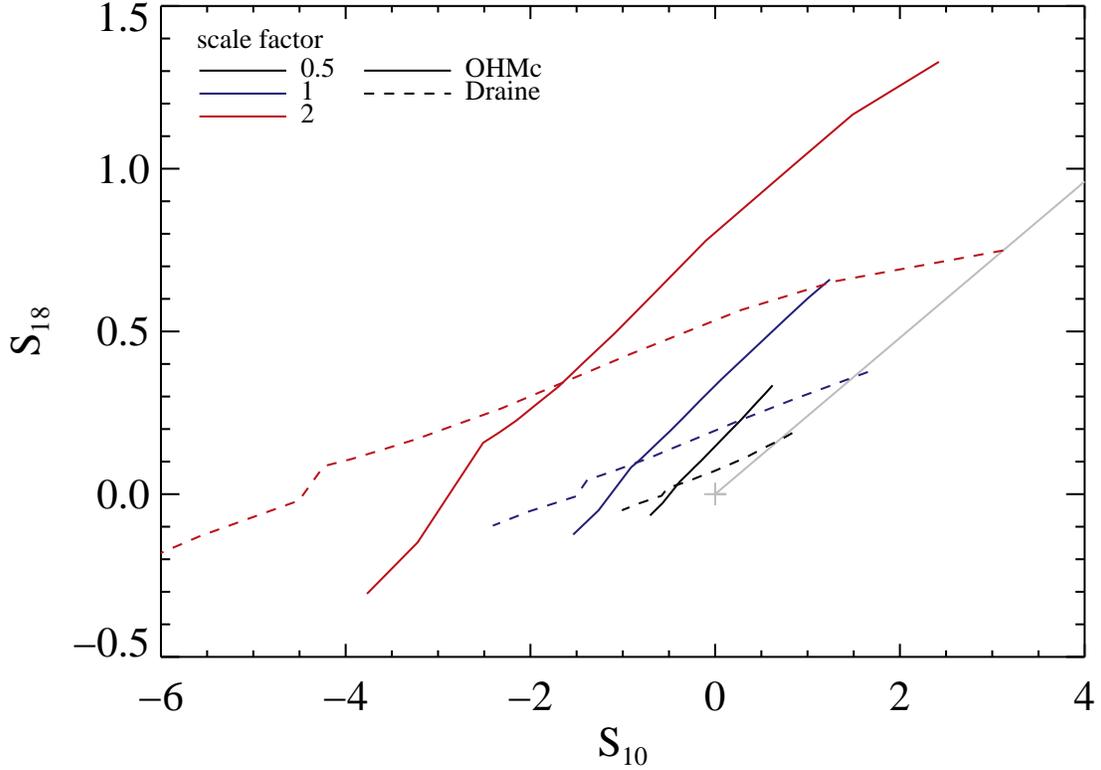}}
\caption{\label{fig:scale}
Effect of artificially scaling feature cross sections with respect to
continuum (described in \S\ref{subsec:artdust}); 
the ratio of 10 and 18 \um{} feature strengths in the dust cross sections 
fully determines the location of the model tracks.
We plot the $p=2$ models, which are independent of $Y$, for $\tau_V = 0$ to 400.
All dust with a given $S^c_{18}/S^c_{10}$ ratio produces
model tracks that have the same slope.
All these tracks start along a line that 
originates at ($S_{10}$, $S_{18}$) = (0, 0)
and has a slope equal to the $S^c_{18}/S^c_{10}$ ratio.
We show this locus for the Draine dust with a light gray line. 
The dust properties cannot be modified with a simple scaling to
match observations of large feature strengths. 
}
\end{figure}

Figure \ref{fig:scale} shows that 
the consequence of this modification is to shift the curves up and down
without changing  their slopes.
We plot
 the results of scaling the Draine silicates
by logarithmic factors of two in the $p=2$ 
models.  We choose this particular density profile  because
the results are independent of
$Y$, yielding a one-parameter family of tracks. 
Density laws $p = 0$ or 1, for 
which shell thickness is important, 
exhibit similar behavior.
All displayed models have the same value of  $S^c_{18}/S^c_{10}$. 
Therefore, the
initial points ($\tau_V = 0$)  of all tracks
lie nearly on the same line that 
originates at the  featureless dust point 
$(S_{10},\ S_{18}) = (0,\ 0)$ 
and has a slope equal to the common $S^c_{18}/S^c_{10}$ ratio.
As is evident in the 
figure,
scaled dust properties 
 also produce 
tracks that have nearly the same slope.
That is, 
{\em the initial point fully determines the slope of the smooth shell tracks.}
Considering an arbitrary line drawn from the featureless point (0, 0), 
all tracks that originate along this line will  have roughly the same slope.

The comparable calculation for the OHMc dust is also plotted,
and its unscaled curve (solid blue line)
serves to indicate the initial trend of the observed strengths.
Evidently, the Draine dust cannot be artificially rescaled to
resemble the OHMc results. 
Having already considered a great range of dust geometries, we conclude that
no scaling can be applied to the Draine or OHMw dust to steepen the strength 
trends of the continuous shell models and  successfully match the observations
at large strength.  The data fundamentally require a higher
ratio of $S^c_{18}/S^c_{10}$, which the OHMc dust provides.
The low  $S^c_{18}/S^c_{10}$ values of the \citeauthor{Chi06} extinction laws
imply that these dust models, too, 
cannot produce tracks that match the observations, in spite
of their relatively high values of $\tau_{sil,18}/\tau_{sil,10}$. 
The reason the \citeauthor{Chi06} dust models couple
a high extinction ratio with a low strength ratio 
is their relatively higher continuum 
around the 18\um\ feature.  Reducing the relative extinction
at long wavelengths would lower the continuum and
thereby increase the relative strength of the 18\um\ feature.

\section{Geometric Families and Optical Classification\label{sec:fam}}

The two silicate strengths together empirically separate the present
ULIRG sample 
into two families
with a discernible gap, 
 as is evident in 
Figures \ref{fig:s18v10ohmc} through \ref{fig:s18v10d}.
ULIRGs in  the first group are deeply absorbed, 
 exhibiting $S_{10}<-1$ and 
 comparably deep 18\um{} absorption.
ULIRGs belonging to the second group have shallow absorption
or emission, with
 $S_{10}\ge -1$ and  commensurate
 18\um{} behavior.  
Theoretically, we account for these two families
in terms of their dust distributions: 
sources in the first group  
are embedded in  smooth dust distributions,
while  those in the second  have
clumpy distributions.
As we noted in \S\ref{subsec:ratio}, the
clumpy tracks populate a distinct region of the feature-feature
diagram that is 
inaccessible to the smooth shell models.
We employed the generic spherical geometry
to minimize the number of free parameters of the models,
but the results remain 
valid even if the sphere is stretched or distorted,
provided the axial ratios do not become extreme \citep{Vin03}.

The first group shows deep absorption, requiring smooth shells, though
ruling out the most concentrated ($p=2$) density distributions.
Members of the second (shallow absorption) group do not lie on the low-$\tau_V$
regime of the continuous shell models.  Instead, they 
are located in the realm of
fragmented shells where
the total optical depth at $V$ is greater than $10 N_0$, and
$N_0 > 1$.
Despite the small magnitude of $S_{10}$ within this group of 
ULIRGs, the average total optical depth of the reprocessing material
must be large, 
although the dust need not be located along the line of sight.
(Indeed, the observation of silicate emission 
requires an unobscured line of sight to some heated cloud faces.)
While the measurement of $S_{10}$ alone does not discriminate between
the smooth  and clumpy distributions, the combination of both
$S_{10}$ and  $S_{18}$ together is diagnostically powerful.

\begin{figure}[htb!]
\centerline{\includegraphics[width=6in]{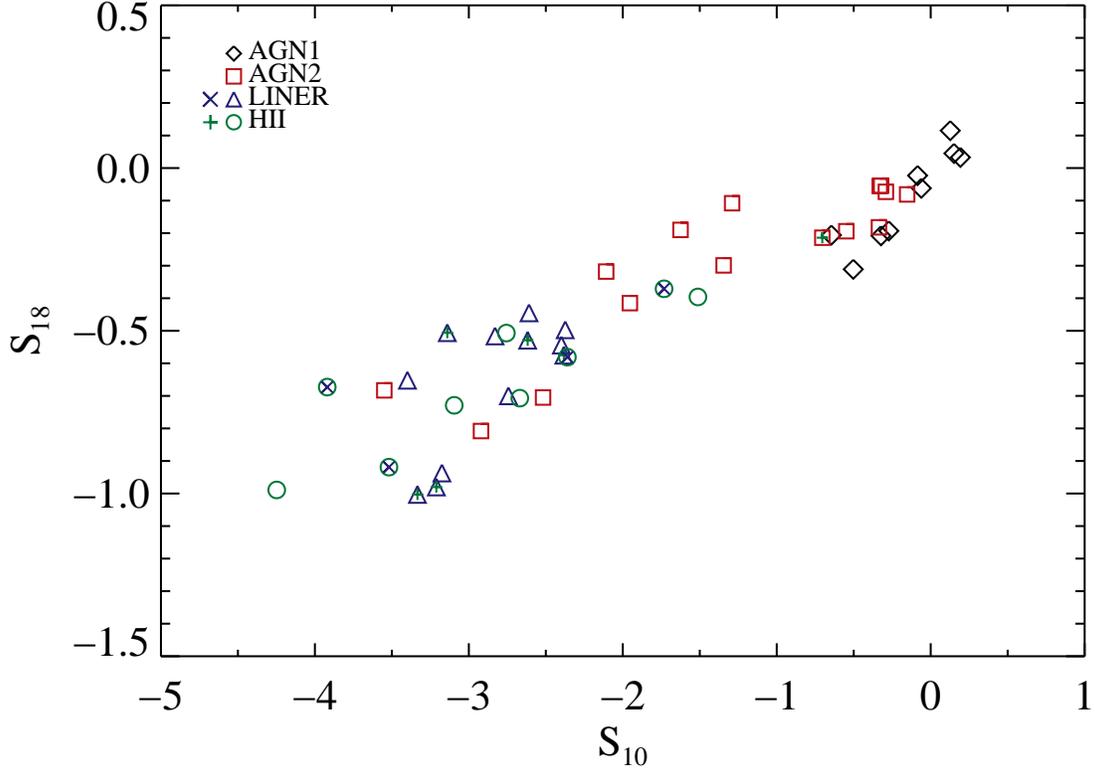}}
\caption{\label{fig:opt}
Measured silicate strengths of ULIRGs coded with optical classifications.
Optical emission line ratios are characteristic of \ion{H}{2} regions,
LINERs, or AGNs.  The AGN spectra are further identified based on
the presence (AGN1) or absence (AGN2) of spectrally broad permitted lines.
Primary or average classifications are plotted with open symbols.
If one diagnostic line ratio or alternate analysis
yields a different classification, 
this secondary identification is also plotted  (\textit{skeletal symbols}).
The LINER and \ion{H}{2} ULIRGs generally show deep absorption, characteristic
of sources embedded in geometrically and optically thick material.
All the AGN1 are members of the clumpy/slab dust distribution family.
Their surrounding optically thick dusty tori are viewed face-on, producing the MIR 
emission while leaving the central engines unobscured. 
The spherical clumpy models
faithfully represent the region of the diagram
that clumpy torus emission occupies (see text for details).
The AGN2 ULIRGs may belong to either family.
}
\end{figure}

The  optical classifications of the weak-PAH ULIRGs we study are related to these two
geometric families.  
Figure \ref{fig:opt} again shows the feature-feature diagram,
adding identification  of the galaxy nuclei
as characteristic of AGNs, LINERs, or \ion{H}{2} regions (listed  in Table \ref{tab:obsv}),
using
optical emission line ratios from the literature.
The clumpy group contains AGNs exclusively, while
the optical LINERs and \ion{H}{2} galaxies 
are all members of the deeply-embedded smooth shell group.
Starting with the latter group, 
the source responsible for the MIR emission from these deeply embedded
ULIRGs
cannot be located in the region that produces
the detected optical lines
because large quantities of dust hide the embedded source
from direct view.  
Conversely, the regions that produce the
optical emission lines  cannot contribute significantly to the MIR
spectrum, because otherwise they would fill the silicate absorption troughs.
For example, 
additional MIR radiation
cannot be present at greater than the 
5\% level in order to measure
 $S_{10} = -3$.
The absence of optically LINER- and \ion{H}{2}-like sources
with $S_{10}>-1$  in our sample 
suggests that 
the IR  emission and optical lines always emanate from different regions
in such ULIRGs.
Thus, we cannot draw any firm conclusions about the embedded sources' 
underlying nature as AGNs or starburst nuclei.
Embedding both an accreting active nucleus and the 
 high ionization line-emitting regions could result in 
optical line ratios that are not typical of AGNs, even though a buried AGN is present.
Independent of whether an AGN or star formation is responsible for
the galaxy's  large and deeply obscured luminosity (and the silicate strengths),
less-buried star formation produces the 
\ion{H}{2} spectra that are detectable at optical wavelengths,
as others have suggested \citep{Ima06,Ima07}.

The ULIRG sample includes objects that are optically identified as
AGNs.  Axial symmetry is the foundation of the standard AGN
unification scheme, with a toroidal dust distribution, whereas the
models we present here employ spherical symmetry.  However, the clumpy
distributions preserve the essence of standard unification, with the
stochastic variations of the clump distribution resulting in
variations of viewing geometry.  In the standard model, both the
central engine and the broad emission line region are obscured from
view when the line of sight passes through the torus, and these are
designated ``type 2'' AGNs.  In these obscured cases, spherical models
yield the same results as equatorial views through a torus, whether
smooth or clumpy, producing 10 and 18\um\ silicate features in
absorption.  On the other hand, when the central engine is viewed
along the symmetry axis, the AGN is unobscured, broad emission lines
are observed, and the AGN is classified as ``type 1.''  Even in this
unobscured case, the observed MIR radiation originates in the
surrounding optically thick torus. Because the toroidal geometry
allows in type 1 viewing a direct line of sight to the heated dust
surface, the silicate features appear in emission
\citep[\textit{e.g.},][]{Pie92}.  Smooth spherical dust configurations
are inapplicable in this case because they offer no unobscured lines
of sight to the central engine.  However, clumpy spherical models do
faithfully reproduce the behavior of the silicate features for both
type 1 and type 2 AGNs; the detailed clumpy torus calculations of
\citet{Nen02,Nen07} produce spectra that are quite similar to those of
the clumpy spheroidal dust distributions we present here.
Fundamentally, {\em clumpy spherical models are valid because they
allow direct views of the AGN and heated cloud surfaces from some
select lines of sight that by chance avoid any obscuring clumps}; the
only significant difference between the two morphologies is that the
fraction of unobscured lines of sight to the central engine increases
when the clump distribution becomes toroidal.  The spherical models
offer the advantage of minimizing the number of free parameters while
retaining the essential ones that govern the behavior of the silicate
features.  The regions of the feature-feature diagrams that these models
occupy are accessible only to clumpy models, irrespective of their
geometry.

In contrast with the deeply embedded ULIRGs,
all sources in the region of clumpy tracks exhibit characteristic 
AGN optical emission, with permitted lines
that are either broad (type 1) or narrow
(type 2).
The location of both type 1 and type 2 AGN ULIRGs
in the clumpy geometric group argues that the
obscuring 
torus itself is clumpy, as first proposed by \citet{Kro88} 
and recently verified for the Circinus AGN
by MIR interferometric observations \citep{Tri07}.
The distribution of AGN types 1 and 2 in the 
feature-feature diagram supports the AGN clumpy 
torus  unification scheme.
All type 1 AGNs are located in the clumpy region.
Indeed, each of these sources offers an unobscured line of sight to the
AGN, 
so the only emitting dust is the torus, which is clumpy.
In contrast, type 2 AGNs are located in both the clumpy
and deeply absorbed regions of the diagram.
AGNs in the former group represent the direct type 2 analogs of the type 1 sources.
The deep absorption features cannot be explained by
the clumpy torus but instead arise naturally from 
additional obscuration by cold foreground dust.
The 
fact that this additional dust is required only in
some type 2, but no type 1, AGNs
agrees with the large extent---hundreds of
pc---of the narrow line region.
Blocking the central engine and the broad line region 
produces a type 2 AGN, irrespective of the  torus orientation. 
A dusty filament  a few tens of pc wide can hide the AGN central engine
without obscuring the narrow line region.
If this dust is not heated by the central source,
it will act as a cold screen  and produce 
an absorption feature.
The depth of the absorption is limited only by the
10\um\ emission of the screen itself, so 
deep absorption can occur when the 
AGN radiation does not heat the dust and the
screen remains cold ($\lesssim 50$K).
For example,
the deepest AGN feature we observe is $S_{10} \approx -3 $,
which 
a single cold molecular cloud with
$\tau_V \approx 50$ could produce.

In principle, a cold screen could account for the deep silicate
absorption in {\em all} these galaxies.  However, in order to remain cold, such a
screen cannot simultaneously reprocess the intrinsic radiation to
emerge at the enormous IR luminosities that identify ULIRGs.  Thus,
the cold screen
must be invoked in addition
to the reprocessing, large optical depth 
 dusty component.
Furthermore, the screen must always fully cover the primary reprocessing
dust along the line of sight.
Such screens  present  a contrived solution for ULIRGs generally,
requiring two separate dust regions 
and correlating the presence of these screens along the line
of sight with the large IR luminosity of ULIRGs.
Moreover, we do not observe any unscreened counterparts of the
optically LINER- and \ion{H}{2}-like sources.
Instead, we find the single entity of embedding
dust to account more plausibly for the total IR characteristics of
deeply absorbed ULIRGs.

We caution that this ULIRG sample is neither complete nor unbiased, 
selected for small $EW_{6.2}$ (less than 0.1).
We cannot draw general conclusions about the distribution of ULIRG properties,
and we emphasize  that the present
 comparison is with only the optical classification, 
not the total evidence for an AGN or starburst nature of the dominant
underlying energy source.
However, 
having already eliminated those ULIRGs that exhibit evidence 
for strong star formation in the MIR from this analysis,
we find a range of optical characteristics
even within this restricted subsample.

Classifying a variety of galaxies on the basis of $S_{10}$ and $EW_{6.2}$, 
\citet{Spo07} identify two distinct branches in the relationship between
these measurements.  They suggest that characteristically 
different dust distributions---either clumpy or smooth---produce 
these two distinct branches, and our results 
validate this conjecture.
While we consider only the small $EW_{6.2}$
subsample of ULIRGs here, it includes members of 
both branches,
 selected where the two branches are
most widely separated.
The weak $S_{10}$ ULIRGs are members of the Spoon et al. (2007)\nocite{Spo07}
``Class 1,'' which they associate with clumpy obscuration.  
Although the weak $S_{10}$ alone could be consistent with either
the smooth or clumpy geometry, 
Figure \ref{fig:s18v10ohmc} demonstrates that
measurements of  $S_{10}$ and $S_{18}$ together
eliminate the possibility of smooth geometry.
The deeply absorbed ULIRGs of the present work are located
along the branch that \citet{Spo07} associate with smooth
dust distributions.  
Indeed, for this group 
both silicate measurements agree 
quantitatively with the
computational results for smooth obscuration, supporting 
the interpretation of \citet{Spo07}.
In addition, we note that \citet{Des07}
include optical identifications with this
 $S_{10}$ and $EW_{6.2}$ classification. 
They find some separation of
optical types among the low-$EW_{6.2}$ objects we study here, 
in agreement with our results based on optical classification
and both silicate feature strengths.

\citet{Ima07} use measurements similar to the silicate strength as
a qualitative indicator of the geometry of the central regions of ULIRGs.
Although they determine the continuum with a local power law fit,
they identify deep absorption  ($S_{10} < -2$)
with an embedded central source, in agreement with this work and with
\citet{Lev07}.
\citet{Ima07} separately consider the strength ratio; 
they  associate a small value of  $S_{18}/S_{10}$ with an embedded central source
and expect a larger ratio when the heating sources are distributed (as in a starburst).
However, our  radiative transfer calculations show that
without considering the strengths of the individual features,
the strength ratio alone is insufficient to determine the dust geometry.
Central heating sources located behind 
low optical depth continuous media,
inhomogeneous distributions, and even foreground screens can produce
low values of  $S_{18}/S_{10}$.
Instead, the  strength ratio depends primarily on dust chemistry, and
both $S_{18}$ and $S_{10}$ are necessary to 
diagnose  dust geometry.

\section{Conclusions}

The large luminosity of ULIRGs requires that
these galaxies possess both an underlying source that is more
energetic than star formation within ordinary galaxies and
dust to reprocess the intrinsic radiation to long wavelengths.
Observations in the IR offer the obvious advantage of 
directly detecting the light that emerges.
Although they suffer
 the disadvantage that dust reprocessing erases the
detailed signature of the nature of the energy source,
as we have shown quantitatively, 
MIR spectroscopy 
does directly probe the dust itself.
Analysis of the silicate features in particular
provides useful diagnostics of the geometric distribution of
dust and its optical properties, which ultimately depend on
its mineralogy.

Comparing observations of ULIRGs with
several models of dust 
chemistry, we find that 
numerical simulations of radiative transfer
 employing the observationally-motivated OHMc dust properties 
 match the data well, and this dust is required 
when deep absorption is measured. 
The key difference between this model and others
that are frequently employed is the relatively high 
silicate absorption peak near 18\um{} relative to that near
10\um; this strength ratio is around 0.5 in the OHMc dust.
These properties are physically-motivated, based on 
laboratory measurements of amorphous silicates.
Including the effects of grain oxidation results in 
the higher ratio of 18/10\um\ cross section, which is
characteristic of bronzite
\citep[and references therein]{Oss92}.
This dust chemistry is not peculiar to the ULIRGs or to the
nuclei of galaxies, but also
fits observations within the Galaxy and Magellanic Clouds.
Moreover, local observations where the silicates appear in
emission rule out the OHMw and  Draine dust at low optical depths. 

Quantifying the strength of the silicate features requires
determining  the underlying continuum accurately.
We demonstrate that including 
measurements of the continuum at intermediate wavelengths
between the features yields physically consistent results,
and we recommend specific modifications to the idealized
method when observational complications, such as photoionized
emission or ice absorption bands, are present.
We emphasize that 
because the same dust is responsible for both the feature and continuum,
the apparent optical depth of silicate absorption
does not reveal the line of sight optical depth,
independent of the continuum-fitting procedure.

We identify two distinct families in the ULIRG measurements,
which is a consequence of fundamental differences in obscuring geometry.
One class 
shows deep absorption ($S_{10}< -1$).
These energetic sources
must be deeply embedded in a continuous medium that is 
geometrically and optically thick ($\tau_V > 100$). 
The other class exhibits weak silicate features, which may
appear in absorption or emission.
Either a 
geometrically thin slab
or a clumpy  dusty medium can account for these observations,
but they do not represent the low optical depth realm of the
continuous geometry.
In addition, we find  that
these two ULIRG families of geometry are related to the 
optical identification of the galaxy nuclei.
All members of the second (low-strength) class are AGNs,
while ULIRGs having spectra characteristic of LINERs and \ion{H}{2} regions
systematically exhibit the deepest absorption.
Among the deeply embedded sources, 
the MIR emission and optical lines always emanate from different regions.
Optically-identified LINERs may be a heterogeneous class,
but the consistent finding 
of \ion{H}{2} ULIRGs
only in the deeply embedded group 
leads us to speculate that
two different regions dominate the IR and 
optical emission line signatures in 
all such optical \ion{H}{2} ULIRGs.

In spite of the general degeneracy of IR SEDs \citep[\textit{e.g.},][]{Vin03}, 
 extreme absorption of the 10\um{} feature alone
proves to be a strong  indicator of
smooth (as opposed to clumpy)  dust distributions \citep{Lev07}.
Here we find that
the combination of
10 and 18\um{} silicate features together 
constrains the geometry 
even more powerfully  
in a wider range of  situations.
Furthermore, the analysis tool of the
feature-feature diagram we introduce here
provides a strong diagnostic of the
dust chemistry.
The optical properties fix the absolute zero points
and the slopes of the 
linear trends of silicate strength that variations of the
model parameters can produce. 
However, 
because analysis based on this diagram  is sensitive only
to the feature strengths in the absorption cross section,
it does not probe feature shape, peak wavelength, or
properties such as mineral composition directly.
We conclude that of the available models, the OHMc dust 
best describes cosmic silicate feature
strengths, both in ULIRGs and local sources. 

\acknowledgements
We thank G. Sloan for stimulating discussions about Galactic dust chemistry 
and 
J. Bernard-Salas, V. Lebouteiller, and D. Whelan for communicating
results and providing reduced spectra in advance of publications.
This work is based in part on observations made with the Spitzer Space Telescope
and has made use of the NASA/IPAC Extragalactic Database,
both of which are
 operated by the Jet Propulsion Laboratory, California Institute of Technology, under contracts with NASA.
NAL acknowledges work supported by the NSF 
under Grant No. 0237291.  ME acknowledges support from
NSF AST-0507421 and NASA NNG05GC38G.

{\it Facility:} \facility{Spitzer (IRS)}

\clearpage
\begin{deluxetable}{lrrrrcl}
\tabletypesize{\footnotesize}
\tablewidth{0pt}
\tablecaption{MIR Measurements and Optical Classification of the ULIRG Sample\label{tab:obsv}}
\tablehead{
\colhead{Galaxy}
&\colhead{$S_{10}$}
&\colhead{$\lambda_{10}$}
&\colhead{$S_{18}$}
&\colhead{$\lambda_{18}$}
&\colhead{Optical Class\tablenotemark{a}}
&\colhead{Optical Reference}\\
}
\startdata
  $00091-0738$    &-3.5 &  9.4 & -0.92 & 18.0 &   H, L & 1 \\ 
  F$00183-7111$    &-2.8 &  9.9 & -0.52 & 18.3 &   L    & 2 \\
  $00188-0856$    &-2.6 &  9.8 & -0.44 & 18.0 &   L    & 1 \\ 
  $00275-2859$    &-0.3 & 10.2 & -0.21 & 18.0 &   A1   & 3 \\ 
  $00397-1312$    &-2.8 &  9.8 & -0.51 & 17.9 &   H    & 1 \\ 
  $00406-3127$    &-2.0 &  9.5 & -0.41 & 17.8 &   A2   & 4 \\ 
  $01003-2238$    &-0.7 &  9.8 & -0.21 & 18.1 &   A2, H & 5, 1 \\ 
$01166-0844$SE    &-3.1 &  9.6 & -0.73 & 17.4 &   H    & 1 \\ 
  $01298-0744$    &-4.2 &  9.9 & -0.99 & 18.0 &   H  & 1 \\ 
  FF J$0139+0115$    &-2.1 &  9.8 & -0.32 & 17.5 &  A2 & 6 \\ 
  $02054+0835$    &-0.1 & 10.8 & -0.02 & 17.9 &   A1   & 7 \\ 
  $03158+4227$    &-3.5 & 10.0 & -0.68 & 17.6 &   A2   & 8  \\
  $05189-2524$    &-0.3 &  9.8 & -0.23 & 18.2 &   A2   & 9 \\ 
  $06035-7102$    &-1.5 &  9.9 & -0.40 & 18.5 &   H    & 10 \\ 
  $06361-6217$    &-2.4 &  9.8 & -0.54 & 17.7 &   L    & 11 \\ 
  $07598+6508$    & 0.2 &  9.6 &  0.05 & 17.2 &   A1   & 12 \\ 
$08572+3915$NW    &-3.9 &  9.9 & -0.67 & 17.8 &   H, L & 13, 14\\ 
  $10091+4704$    &-3.3 &  9.9 & -1.00 & 18.3 &   L, H & 14  \\ 
  $10378+1109$    &-2.4 &  9.9 & -0.50 & 17.9 &   L    & 1   \\ 
  $11095-0238$    &-3.4 &  9.9 & -0.65 & 18.2 &   L    & 10   \\ 
  $11582+3020$    &-3.2 & 10.0 & -0.94 & 18.2 &   L    & 14  \\ 
  $12032+1707$    &-2.7 &  9.8 & -0.70 & 18.3 &   L    & 1   \\ 
  $12071-0444$    &-1.3 &  9.9 & -0.30 & 18.2 &   A2   & 12  \\ 
$12127-1412$NE    &-2.4 &  9.8 & -0.57 & 17.8 &   L, H & 1 \\ 
  $12514+1027$    &-1.6 &  9.8 & -0.19 & 18.0 &   A2   & 15 \\ 
  $13218+0552$    &-0.5 &  9.9 & -0.31 & 18.2 &   A1   & 16 \\ 
  $13451+1232$    &-0.3 &  9.8 & -0.05 & 18.4 &   A2   & 12 \\ 
  $14070+0525$    &-2.9 & 10.0 & -0.81 & 18.0 &   A2   & 14 \\ 
  F$14548+3349$    &-2.5 & 10.6 & -0.70 & 17.9 &   A2   & 17 \\ 
  $15225+2350$    &-2.4 &  9.9 & -0.58 & 17.8 &   H, L & 1  \\ 
  $15250+3609$    &-3.1 &  9.8 & -0.51 & 18.0 &   L, H & 18 \\ 
  $15462-0450$    &-0.3 & 10.3 & -0.19 & 18.4 &   A1   & 10  \\ 
  $16090-0139$    &-2.6 &  9.9 & -0.53 & 18.1 &   L, H & 14 \\ 
  $17044+6720$    &-1.7 &  9.9 & -0.37 & 18.0 &   H, L & 19, 14 \\ 
  $17179+5444$    &-0.2 &  9.2 & -0.08 & 18.0 &   A2   & 1  \\ 
 $19254-7245$S    &-1.3 &  9.4 & -0.11 & 18.3 &   A2   & 10 \\ 
  $20037-1547$    &-0.1 & 10.1 & -0.06 & 18.8 &   A1   & 7 \\ 
  $20100-4156$    &-2.7 & 10.1 & -0.71 & 18.1 &   H    & 10 \\ 
  $23060+0505$    &-0.3 &  9.7 & -0.07 & 17.8 &   A2   & 20 \\ 
  $23129+2548$    &-3.2 & 10.0 & -0.98 & 18.0 &   L, H &  1 \\ 
  $23498+2423$    &-0.5 &  9.8 & -0.19 & 18.3 &   A2   &  1 \\ 
  F$23529-2199$    &-0.3 &  9.3 & -0.05 & 18.4 &   A2   & 21 \\ 
          3C273   & 0.1 & 10.9 &  0.12 & 17.5 &   A1   & \nodata    \\  
   Mrk 1014       & 0.2 &  9.5 &  0.03 & 17.4 &   A1   & 22 \\ 
     Mrk 231      &-0.6 &  9.8 & -0.21 & 17.6 &   A1   & 9 \\ 
     Mrk 463E      &-0.4 &  9.8 &  0.09 & 18.3 &   A2   & 22 \\
\enddata                                                                                     
\tablenotetext{a}{Optical classification: A1 = broad-line AGN; A2 = narrow-line AGN; L = LINER; H = H \textsc{ii} galaxy.  Multiple classes are listed when various diagnostic diagrams or alternative analysis 
yields different results, with the  primary classification listed first.}
\tablerefs{(1) Veilleux et al. 1999; 
(2) Armus et al. 1989; 
(3) Vader \& Simon 1987; 
(4) Allen et al. 1991; 
(5) Farrah et al. 2005; 
(6) Stanford et al. 2000;
(7) Lawrence et al. 1999; 
(8) Risaliti et al. 2000;
(9) Sanders et al. 1988a;
(10) Duc et al. 1997;
(11) Keel et al. 2005;
(12) Sanders et al. 1988b;
(13) Arribas et al. 2000;
(14) Kim et al. 1998;
(15) Cutri et al. private communication, cited in Rowan-Robinson 2000;
(16) Low et al. 1988; 
(17) Rupke et al. 2005;
(18) Veilleux et al. 1995;
(19) deGrijp et al. 1992;
(20) Frogel et al. 1989;
(21) Clements et al. 1996;
(22) Mazzarella \& Balzano 1986.
}
\end{deluxetable}

\begin{deluxetable}{lcccc}
\tablewidth{0pt}
\tablecaption{Silicate Properties of Model Dust\label{tab:dust}}
\tablehead{
\colhead{Dust}
&\colhead{$S^c_{10}$}
&\colhead{$S^c_{18}$}
&\colhead{$S^c_{18}/S^c_{10}$}
&\colhead{$\tau_{sil, 18}/\tau_{sil, 10}$}\\
}
\startdata
OHMc         &  1.26 &    0.67 &   0.54 &  0.49 \\
OHMw         &  1.26 &    0.43 &   0.34 &  0.39 \\
Draine       &  1.69 &    0.38 &   0.23 &  0.37 \\
\enddata
\tablecomments{Silicate strengths measured in dust cross sections,
$S^c$, following Equation \ref{eqn:sdef}, as applied to
cross sections plotted in Figure \ref{fig:cross}.
The ratio $\tau_{sil, 18}/\tau_{sil, 10}$
is calculated from the  total cross sections at the feature peaks.
}
\end{deluxetable}

\clearpage

\clearpage

\clearpage

\clearpage

\clearpage

\end{document}